\documentclass[preprintnumbers,nofootinbib,aps,prd,floatfix,twocolumn]{revtex4}

\pdfoutput=1
\usepackage{graphicx}
\usepackage{latexsym}
\usepackage{amsfonts}
\usepackage{amssymb}
\usepackage{amsmath}
\usepackage{slashed}
\usepackage{mathtools}
\usepackage{hyperref}
\usepackage{url}
\usepackage{color}
\usepackage{cancel}
\usepackage{multirow,array}
\usepackage{float}
\usepackage{subfigure}
 
\usepackage[normalem]{ulem}

\begin{document}
\title{Measuring Galactic Dark Matter through Unsupervised Machine Learning}

\author{Matthew R.~Buckley}
\affiliation{Department of Physics and Astronomy, Rutgers University, Piscataway, NJ 08854, USA}

\author{Sung Hak Lim}
\affiliation{Department of Physics and Astronomy, Rutgers University, Piscataway, NJ 08854, USA}

\author{Eric Putney}
\affiliation{Department of Physics and Astronomy, Rutgers University, Piscataway, NJ 08854, USA}

\author{David Shih}
\affiliation{Department of Physics and Astronomy, Rutgers University, Piscataway, NJ 08854, USA}

\begin{abstract}
Measuring the density profile of dark matter in the Solar neighborhood has important implications for both dark matter theory and experiment. In this work, we apply autoregressive flows to stars from a realistic simulation of a Milky Way-type galaxy to learn -- in an unsupervised way -- the stellar phase space density and its derivatives. With these as inputs, and under the assumption of dynamic equilibrium, the gravitational acceleration field and mass density can be calculated directly from the Boltzmann Equation without the need to assume either cylindrical symmetry or specific functional forms for the galaxy's mass density. We demonstrate our approach can accurately reconstruct the mass density and acceleration profiles of the simulated galaxy, even in the presence of {\it Gaia}-like errors in the kinematic measurements.
\end{abstract}

\maketitle

\section{Introduction}

Dark matter is definitive evidence for physics beyond the Standard Model. No known particle has the necessary properties to resolve the gravitational anomalies which arise in spiral galaxies \cite{1980ApJ...238..471R,1939LicOB..19...41B,Salucci:2018hqu}, galaxy clusters \cite{2011ARA&A..49..409A,1933AcHPh...6..110Z}, early Universe cosmology \cite{Planck:2018vyg}, and gravitational lensing \cite{Clowe:2003tk} if non-relativistic energy density was only sourced by the visible baryonic material. Despite an extensive experimental particle physics program across direct detection, indirect detection, and collider searches, astrophysics remains the only source of definitive positive knowledge about the particle nature of dark matter. Analyzing the gravitational influence of dark matter remains a critical source of new information which informs theoretical ideas about this mysterious substance (see {\it e.g.}, Ref.~\cite{Buckley:2017ijx} for a review).  

With this context in mind, the distribution of dark matter in the Solar neighborhood and more broadly throughout the Milky Way galaxy is of immense interest to the study of the physics of the dark sector. The density of dark matter at the Earth's location enters directly into the scattering rate for direct detection experiments, and the distribution of dark matter within the Galaxy determines the intensity and angular distribution of indirect detection signals of dark matter annihilation and decay at the Galactic Center. Additionally, non-trivial physics within the dark sector -- for example, self-interacting \cite{Spergel:1999mh}, fuzzy \cite{Hu:2000ke}, or dark disk \cite{Fan:2013tia,Fan:2013yva} dark matter -- can modify the density profile of the Galaxy's dark matter halo. Relatedly, measurements of the acceleration field off of the Galactic disk can discriminate between dark matter and modified gravity solutions \cite{Buch:2018qdr}. 

A large literature exists \cite{1984ApJ...276..169B,1984ApJ...287..926B,2006A&A...446..933B,2012ApJ...756...89B,2013ApJ...779..115B,2018MNRAS.473.2288B,Buch:2018qdr,2010JCAP...08..004C,1998A&A...329..920C,2012MNRAS.425.1445G,2020MNRAS.495.4828G,2018A&A...615A..99H,2021MNRAS.508.5468H,2000MNRAS.313..209H,2004MNRAS.352..440H,1922ApJ....55..302K,1989MNRAS.239..605K,1989MNRAS.239..651K,1989MNRAS.239..571K,1991ApJ...367L...9K,2011MNRAS.414.2446M,2012MNRAS.419.2251M,2013MNRAS.433.1411M,2012ApJ...751...30M,2020MNRAS.494.6001N,2021ApJ...916..112N,1932BAN.....6..249O,1960BAN....15...45O,2019A&A...621A..56P,2020A&A...643A..75S,2018PhRvL.121h1101S,2003A&A...399..531S,2018MNRAS.478.1677S,1996MNRAS.282..223S,2019MNRAS.485.3296W,2019A&A...623A..30W,2021A&A...646A..67W,2019MNRAS.482..262W,2013ApJ...772..108Z,2020EPJWC.24004002W} which seeks to measure the dark matter density in the local volume of space via its effect on the statistical properties of a population of visible tracer stars' kinematics. These stars are moving within the Galactic gravitational potential $\Phi(\vec{x})$, sourced by stars, gas, and dark matter. For a population of tracers which are well-mixed into the Galaxy, the phase space density as a function of position and velocity $f(\vec{x},\vec{v})$ respects the collisionless Boltzmann Equation:\footnote{Here, and throughout this paper, Cartesian coordinates are indexed by Latin characters $i,j,k$ and the Einstein summation convention is assumed.}
\begin{equation}
\frac{d f}{d t}+v_i\frac{\partial f}{\partial x_i} = \frac{\partial\Phi}{\partial x_i}\frac{\partial f}{\partial v_i}  \label{eq:boltzmann_mastereq}.
\end{equation}
If the tracer population is in equilibrium, then $\partial f/\partial t$ is zero. There is compelling evidence that the Milky Way local to the Sun is not currently in dynamic equilibrium \cite{2012ApJ...750L..41W,2013MNRAS.436..101W,2018Natur.561..360A}, due to interactions with the Sagittarius stream as well as the Magellanic Clouds. Despite this, when inferring density and acceleration using the Boltzmann equation, equilibrium is typically assumed (see for example Ref.~\cite{2021RPPh...84j4901D}); we will do the same in this work. However, in the absence of equilibrium our proposed direct measurements of $f(\vec{x},\vec{v})$ provide interesting possibilities when combined with other measurements of Galactic acceleration; we will return to this point later in the paper.

Using the Boltzmann equation -- which relates the derivative of the potential (that is, the gravitational acceleration) to derivatives of the full six-dimensional phase space density of a stellar population -- has historically proven difficult. The high dimensionality of phase space implies that any attempt to construct $f$ from data through binning will result in poor statistics of most bins in $\vec{x}$ and $\vec{v}$. These inaccuracies will then propagate and amplify in the calculation of the derivatives. While other methods have been developed (distribution function modeling \cite{2012MNRAS.419.2251M,2013MNRAS.433.1411M,2019A&A...621A..56P,2021MNRAS.508.5468H} or made-to-measure methods \cite{1996MNRAS.282..223S,2018MNRAS.473.2288B}) a common approach to overcome this problem is to take the first moment of the Boltzmann equation and integrate over velocity. With the spatial phase space of the tracers defined as $\nu(\vec{x}) = \int d^3\vec{v} f(\vec{x},\vec{v})$, the resulting Jeans Equation is
\begin{equation}
\overline{v_i v_j} \frac{\partial \nu(\vec{x})}{\partial x_i} + \nu(\vec{x}) \frac{\partial \overline{v_i v_j}}{\partial x_i} = -\nu(\vec{x}) \frac{\partial \Phi}{\partial x_j}, \label{eq:jeans}
\end{equation} 
where the average squared velocity matrix
\begin{equation}
\overline{v_i v_j} \equiv \frac{1}{\nu(\vec{x})} \int d^3\vec{v} f(\vec{x},\vec{v}) v_i v_j  \label{eq:v2def}
\end{equation} 
is position-dependent.  An alternative approach to Jeans modeling is to assume a flexible form of the distribution function $f$, take analytic derivatives, and fit to data. In practice, fitting to data regardless of approach often requires additional assumptions of axisymmetry, small ``tilt'' terms which mix radial and $z$ components (here $z$ is defined as the direction off of the Galactic disk, with the midline set as $z=0$), and/or specific assumptions about the functional forms of the tracer population density. 

Once the accelerations $-\partial\Phi/\partial\vec{x}$ have been determined, the total density can be obtained from the Poisson Equation
\begin{equation}
4\pi G \rho = \nabla^2 \Phi. \label{eq:poisson}
\end{equation}
Subtracting the measured stellar and gas densities then gives the dark matter density. We refer to Refs.~\cite{2018MNRAS.478.1677S,2020A&A...643A..75S,2020MNRAS.494.6001N,2021ApJ...916..112N,2020MNRAS.495.4828G,2018A&A...615A..99H} for recent Jeans analysis of the Milky Way density, Refs.~\cite{Buch:2018qdr,2018PhRvL.121h1101S,2019MNRAS.482..262W,2019A&A...623A..30W,2021A&A...646A..67W,2021MNRAS.508.5468H} for recent distribution function fits, and Refs.~\cite{2014JPhG...41f3101R,2021RPPh...84j4901D} for comprehensive reviews.

Refining assumption-free approaches to determine the Galactic acceleration and density fields is a timely question, given the exquisite kinematic data from the {\it Gaia} Space Telescope \cite{2016Gaia}. {\it Gaia}'s unique capability in stellar measuring proper motion and vast dataset is revolutionizing our understanding of Milky Way structure. In its third early data-release (eDR3) \cite{2021Gaia}, {\it Gaia} has measured the position, parallax, and proper motion of 1.5 billion stars, and can further measure radial motion for 7.2 million; further data releases will only increase the quantity and quality of these data sets.  In this work, we will assume measurements of synthetic data which broadly match the performance of the {\it Gaia} mission as we consider the problem of extracting mass density from a population of tracer stars.

Recently, a new approach to modeling the gravitational acceleration in the Milky Way has been proposed, first in Ref.~\cite{2020arXiv201104673G} and elaborated in Refs.~\cite{2021MNRAS.506.5721A,2021arXiv211207657N}. This method uses the Boltzmann Equation directly, leveraging new machine learning techniques which can accurately estimate high-dimension phase space directly from data. These unsupervised machine learning architectures -- known as normalizing flows (see Ref.~\cite{9089305} for a review) -- learn to transform a simple (typically Gaussian) probability density in a non-physical ``latent'' space into a probability density of an arbitrary dataset in a multi-dimensional space. (For a previous application of normalizing flows to actual {\it Gaia} data, to perform a data-driven, model-independent search for stellar streams, see Ref.~\cite{Shih:2021kbt}.) In the context of the Boltzmann Equation for Galactic kinematics, these networks provide a differentiable function $f(\vec{x},\vec{v})$ from the tracer star population without requiring specific functional form assumptions or necessarily imposing symmetry constraints on the potential.  

Using spherically symmetric mock stellar data, Ref.~\cite{2020arXiv201104673G} modeled the phase-space density distributions using normalizing flows, and then used a second neural network to model the gravitational potential directly, with a loss function that solves the Boltzmann equation. Ref.~\cite{2021MNRAS.506.5721A} also used a neural network to construct the phase-space density terms of the Boltzmann Equation, and then numerically solved for the derivatives $\partial\Phi/\partial\vec{x}$, leveraging the fact that the density-estimating network can be evaluated at multiple $\vec{v}$ values for a single position. Using one million mock stars within 1~kpc of the Sun, drawn from an analytic model of the Milky Way, Ref.~\cite{2021arXiv211207657N} further demonstrated that this approach can reconstruct the gravitational acceleration field with high fidelity, even in the presence of realistic measurement errors.

In this paper, we extend the approach of Refs.~\cite{2020arXiv201104673G,2021MNRAS.506.5721A,2021arXiv211207657N} in two important ways. First, we demonstrate for the first time that the numeric solutions for the derivatives of $\Phi$ from phase-space density estimators can be extended to the second derivative, and thus the total mass density can be extracted directly from the density estimator. Second, we use mock stellar data drawn from fully cosmological simulations \cite{2012ApJ...761...71Z,2012ApJ...758L..23L} with realistic errors of the type expected from the {\it Gaia} Space Telescope \cite{2021Gaia,2016Gaia}, simulating stars up to 3.5~kpc from the Sun. With our approach, and using statistical errors similar to what could be achieved using the {\it Gaia} telescope, we show the Milky Way dark matter acceleration and mass density profile can be reconstructed out to distances of at least $\sim 1.5$~kpc, even in the presence of disequilibria and with datasets much smaller than would be expected from real data.

In Section~\ref{sec:data} we describe the simulated stellar data we will use to train our normalizing flows. The network itself is described in Section~\ref{sec:flows}, where we also include a discussion of error propagation. Section~\ref{sec:boltzmann} contains our method of calculating the acceleration from the tracer dataset, and  Section~\ref{sec:density} describes our method for calculating the mass density. Conclusions and directions for future work are discussed in Section~\ref{sec:conclusions}.

\section{Simulated Stellar Kinematics \label{sec:data}}

In order to test a technique to measure the gravitational potential from the motion of tracer stars in a system where truth-level information is known, we must start with a set of simulated stars which have kinematics that are self-consistent with the underlying gravitational potential. Fortunately, such a system is available from cosmological $N$-body simulations of galaxies. Such simulated galaxies are constructed from a spectrum of initial density perturbations drawn from early Universe cosmology, and then evolved forward in time, allowing individual particles representing collections of stars, gas, and dark matter to interact through gravitational forces (and, in the case of stars and gas, through supernovae feedback, star formation, and other baryonic effects). In this section, we describe the simulated galaxy we adopt as our Milky Way analogue, the construction of the synthetic set of tracer stars, and the addition of realistic measurement errors to this dataset.

\subsection{Milky Way-Like Simulated Galaxy}

\begin{figure*}
\includegraphics[width=2.0\columnwidth]{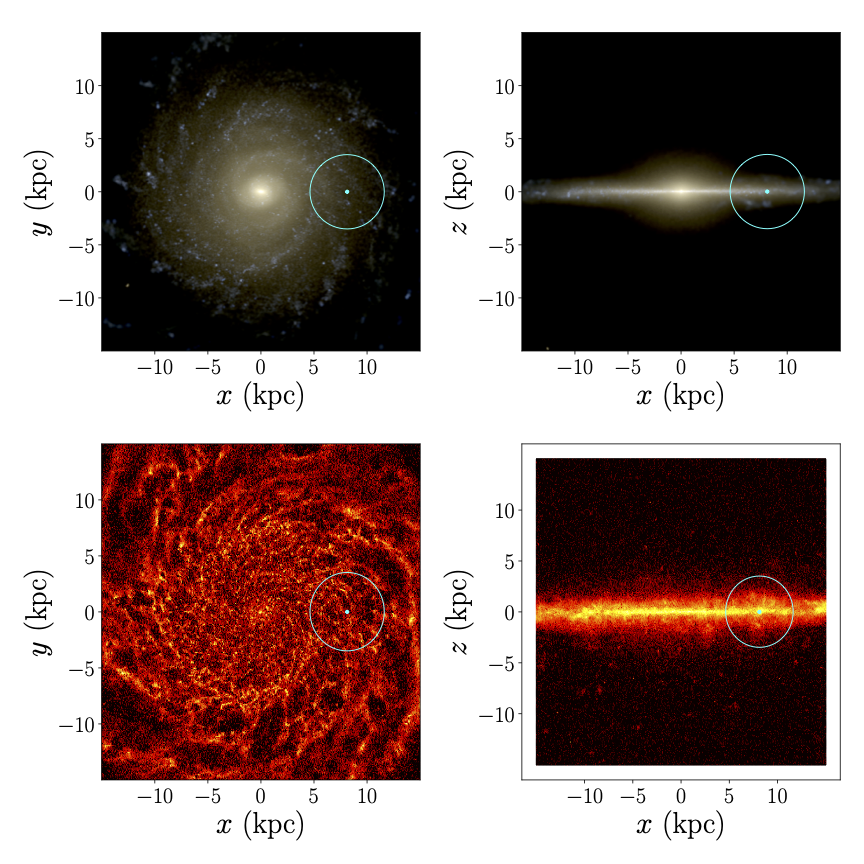}

\caption{Upper Row: Simulated image of the stellar population of \textsc{h277} in the $i$, $u$, and $v$ bands (using \textsc{PynBody} visualization functions), with viewing angle oriented along $+z$ axis (left) and in the plane of the disk (right). The cyan dot indicates the location of our assumed center of observation (the ``Sun'') and the cyan circle shows the 3.5~kpc observation window. Lower Row: the gas distribution (with intensity indicated in a log-scale of color) of \textsc{h277}, oriented as in the upper panels.   
\label{fig:h277_overview}}
\end{figure*}

In this work, we use the simulated galaxy \textsc{h277} as the source for our stellar kinematics dataset \cite{2012ApJ...761...71Z,2012ApJ...758L..23L}.\footnote{The simulation data is available from the $N$-Body Shop Collaboration at \url{https://nbody.shop/data.html}.} This galaxy is the result of a fully cosmologically $N$-body simulation using \textsc{Gasoline}, an $N$-body smoothed particle hydrodynamics code \cite{2004NewA....9..137W}. The galaxy was selected from a dark matter-only initial simulation using a uniform-resolution box with side-length of $50$~Mpc (in comoving units). The region within the box was then resimulated with the same initial conditions at a higher resolution, this time including baryons. The high-resolution simulation has a force resolution of $173$~pc in the region of interest, dark matter $N$-body particle mass of $1.3\times 10^5\, M_\odot$, and high-resolution initial gas particle mass of $2.7 \times 10^4\,M_\odot$, though some low-resolution gas particles with much higher masses are present. The spiral galaxy \textsc{h277} has a mass of $7.95\times 10^{11}\,M_\odot$ within the virial radius of 312~kpc at the redshift zero snapshot used in our analysis, and is the closest Milky Way analogue from the suite of simulated galaxies available from this code \cite{2014ApJ...794..151L}. Figure~\ref{fig:h277_overview} shows the stellar disk and gas distribution of \textsc{h277}.

We interface with the redshift-zero snapshot of the full cosmological simulation using \textsc{PynBody} \cite{pynbody}, and use this code to identify the barycenter of the galaxy's halo and orient the disk into the $x-y$ plane. This defines our ``galactocentric'' coordinate system. We define the positive $z$ axis (which is perpendicular to the disk) so that the net rotation of disk stars moves in the positive $y$ direction in a right-handed coordinate system.

The simulation allows us to access the truth-level acceleration and mass density. The ``true'' acceleration at any position is calculated through direct summation of the gravitational acceleration due to every other particle (stars, gas, and dark matter) in the simulation, with a softening length set by the simulation resolution. 
The ``true'' mass density is a sum of cubic spline kernels centered the location of every star, dark matter, and gas particle in the simulation.
The resolution limit of the simulation ($\epsilon=0.173$~kpc for particles within $\sim 100$~kpc of the \textsc{h277} galactic center) should be kept in mind when comparing these truth-level results with our calculated densities and accelerations, and will require some averaging in order to obtain non-stochastic ``truth-level'' accelerations and mass densities; these smoothing functions are defined in Sections~\ref{sec:boltzmann} and \ref{sec:density}.

The last major merger occurred at redshift $\sim 3$ for \textsc{h277}, suggesting that the galaxy is largely in equilibrium at redshift zero, though we do not enforce equilibrium by hand. In the Milky Way, the spiral arms can correspond to ${\cal O}(20\%)$ mass overdensities at Galactic radii close to the Sun \cite{2016MNRAS.461.2383P}. We find similar deviations in mass density due to the stars and gas in the spiral arms of \textsc{h277}. The passage of these spiral arms cause the phase space density of disk stars to change with time, even if the overall structure of the galaxy is in equilibrium. 
Similarly, though the galactic disk is largely axisymmetric, local deviations can occur and we do not enforce or assume axisymmetry at any point in our neural network analysis.

\subsection{Stellar Tracers}

We rotate the galactocentric Cartesian coordinate system so that the observational center for our analysis (the ``Sun's'' location) is along the positive $x$-axis, and at the same distance from the \textsc{h277} center and moving in the same direction as in the real Milky Way: $\vec{x}_\odot = (8.122,0.0,0.0208)$~kpc and $\vec{v}_\odot = (12.9,245.6,7.78)$~km/s \cite{2010MNRAS.403.1829S}. After these definitions, the Sun within \textsc{h277} is co-rotating with the disk stars, as it is in the Milky Way. 

We select all star particles within 3.5~kpc of the Sun to construct our kinematic dataset. Without measurement errors, this corresponds to 153,714 star particles. In Figure~\ref{fig:h277_overview}, we show the stellar disk of \textsc{h277}, the gas distribution, and our assumed Solar location. With measurement errors (see the next subsection), 160,881 stars appear to be within 3.5~kpc of the Sun. 

In converting the simulation of star particles into a set of mock observations of individual stars, we must recall that the star particles in the simulation have an average mass of $\sim 5800\,M_\odot$. Each of these star particles corresponds to thousands of individual stars, which must populate a region of kinematic phase space around the star particle itself. Converting a galaxy simulation with ${\cal O}(10^6)$ star particles into a realistic collection of ${\cal O}(10^{11})$ stars like the Milky Way requires up-sampling the star particles into their constituent stars, followed by applying realistic observational effects. This has been accomplished for a number of cosmological simulations \cite{ananke,aurigaia}, resulting in mock astronomical observations for surveys such as {\it Gaia}. 

However, there is reason to be concerned that such up-sampling might introduce unphysical errors when our method is applied. Existing methods which are used to calculate the phase space density of the $N$-body star particles\footnote{{\it e.g.}, Gaussian sampling around each star particle, using the codes such as \textsc{EnBid} \cite{2006MNRAS.373.1293S} to estimate the kernel.} -- required for up-sampling -- causes unphysical clumping of up-sampled stars around each particle, which could bias the estimated accelerations and mass densities.

To avoid this issue, in this work we construct our sample of ``stars'' from the simulation star particles in a much simpler way. Rather than upsampling each star particle into many stars, we assume each star particle contains exactly one star which is measured in our dataset. 
We will assume each of these tracer stars is an RR Lyrae variable, allowing for reasonable measurements of the stars' distances from the Sun to be inferred using luminosity measurements, as described in the next subsection. 

The relatively small tracer star dataset, combined with the gas and stars in the spiral arms in the simulation present something of a challenge for our analysis, as small-scale inhomogeneities can disrupt the assumptions of local equilibrium in the regions of the disk with high variations in the gas and star density.  In analyses of the Milky Way, selecting stars based on age and metallicity to have a tracer dataset of halo stars reduces the impact of these local spiral disequilibria on mass density calculations which rely on the Boltzmann Equation. This is because the time-dependent population of disk stars and gas causes a non-zero $df/dt$ for halo stars only though the gravitational interactions between the disk and the halo. However, if we were to use star particle metallicity as a selection criteria to restrict ourselves to halo stars, we would have only ${\cal O}(10^4)$ stars to train our density estimator on. This is far too few to achieve accurate fits. As a result of using tracers including the disk stars, the non-zero $df/dt$ from the spiral arms enters the Boltzmann Equation at leading order. 

 With this in mind, we chose our ``Solar location'' within \textsc{h277} so as to avoid the spiral arms as much as possible (though the churning of the disk stars from the bulge remains an issue on the edge of our observational window closest to the galactic center). We again emphasize that this avoidance of the spiral arms is necessary only because selecting the tracer population to enrich halo stars is not an option given the small dataset, and we do not expect it this to be an issue when this technique is applied to real data. We also note that this issue of local disequilibrium due to spiral arms is not present in simulated galaxies which draw stars from smooth distributions without spiral arms. 
 
It should be noted that, compared to the datasets on which similar acceleration and/or mass densities analyses are typically performed, this dataset is  small. For example, Ref.~\cite{2020MNRAS.494.6001N} has $1.98\times 10^6$ stars within a region of the Milky Way approximately the size of the observation sphere we consider in this paper. Ref.~\cite{2021arXiv211207657N} simulated $10^6$ stars within 1~kpc of the Solar location, to be compared with the $\sim 8,000$ 
stars inside that radius in our dataset. In addition, due to the dominance of disk stars, our dataset is very sparse at higher values of $|z|$. For example, only $\sim 30\%$ of our tracer stars have $|z|>0.5$~kpc, and only $10\%$ have $|z|>1$~kpc, despite these regions being 80\% and 60\% of the total volume of the 3.5~kpc observation sphere. Real data, with presumably a larger number of tracer stars, may provide better fits to the inferred phase-space density than we find in this study.

\subsection{Measurement Error Model \label{subsec:errormodel}}

We wish to simulate realistic measurements of stellar kinematics as obtained from the {\it Gaia} space telescope. To account for measurement errors in a realistic dataset, we add randomized errors to the true angular position, proper motion, distance, and radial velocity (relative to Earth).  As we shall see, the largest contributor to the kinematics errors is the distance. Though {\it Gaia} can measure stellar distances directly through parallax, the errors grow very large for stars even a few kiloparsecs away.

Instead, by selecting RR Lyrae stars, we can instead use their standard luminosity to determine distance. We assume an absolute magnitude of $0.64$ and a Gaussian error with standard deviation of $0.25$ \cite{2019MNRAS.482.3868I}. This corresponds to approximately 10\% errors on the distance for stars in our sample. Errors for angular position, proper motion, and radial (line-of-sight) velocities are obtained from the {\it Gaia} Data Release 3 parameters \cite{2021Gaia} as implemented in the \textsc{PyGaia} software package \cite{pygaia}, and are of ${\cal O}(15~\mu{\rm as})$ for the position, ${\cal O}(20~\mu{\rm as/yr})$ for proper motion, and ${\cal O}(20~{\rm km/s})$ for radial velocity. The expected errors for each of these quantities depends on the apparent magnitude of the star, and this is accounted for in our error generation. Conversion from the galactocentric Cartesian coordinates to the Earth-centered International Celestial Reference System (ICRS) is handled using built-in \textsc{AstroPy} \cite{astropy:2013,astropy:2018} functions, with the appropriate coordinate transformation so that the assumed Solar motion co-rotates with the disk.

As the distance enters in the conversion from the angular position and proper motion back to Cartesian coordinates, these latter sources of errors are completely subdominant to the distance error.  The error in the radial velocity is the second largest single source of error. 


\begin{figure*}[th]
\includegraphics[width=1.6\columnwidth]{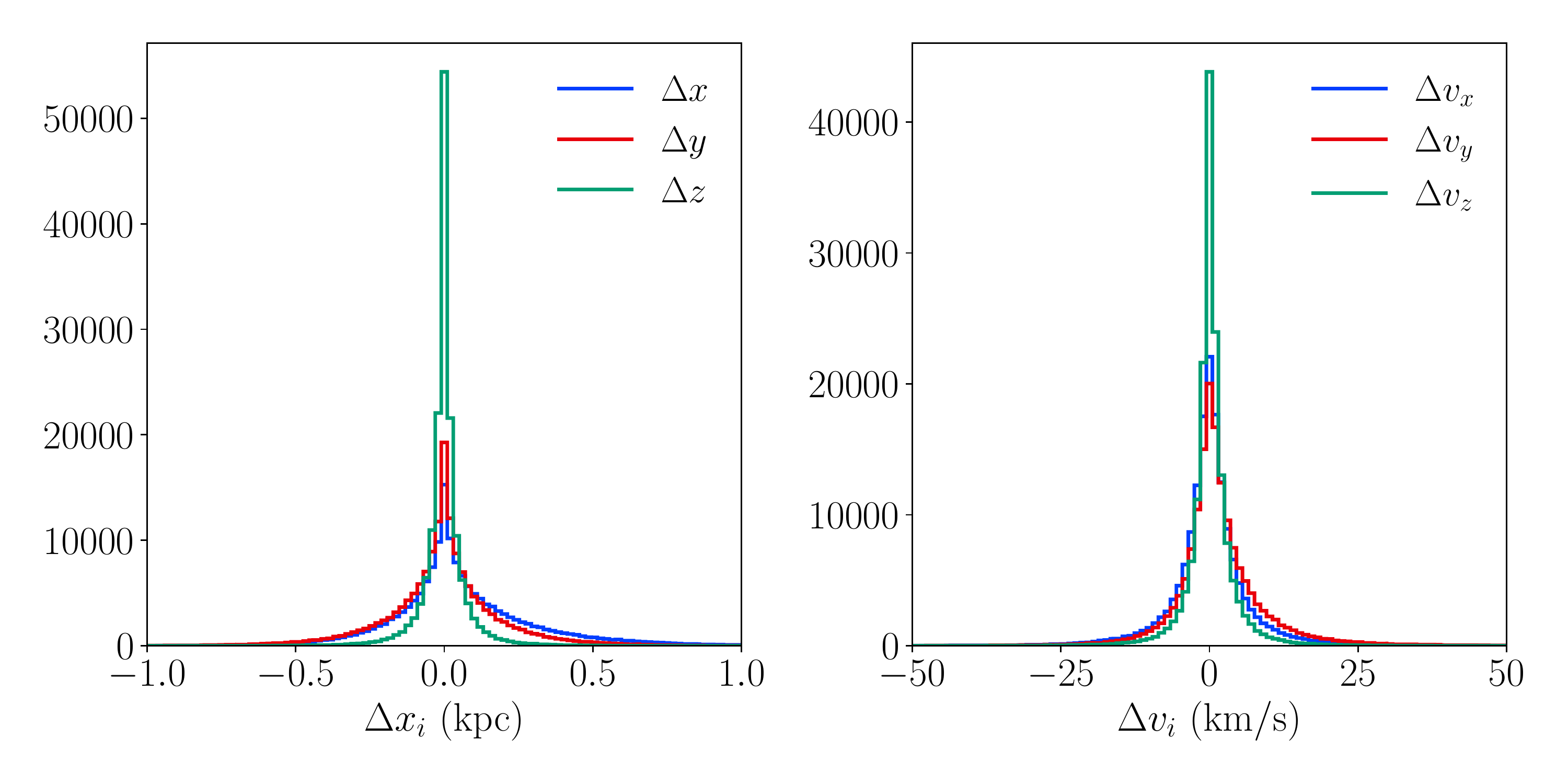}

\caption{Shift in star position (left) and velocity (right) in galactocentric Cartesian coordinates after the application of measurement errors for all stars which are within 3.5~kpc of the Solar position. \label{fig:data_errors}}
\end{figure*}

\begin{figure*}[ht]
\includegraphics[width=2.0\columnwidth]{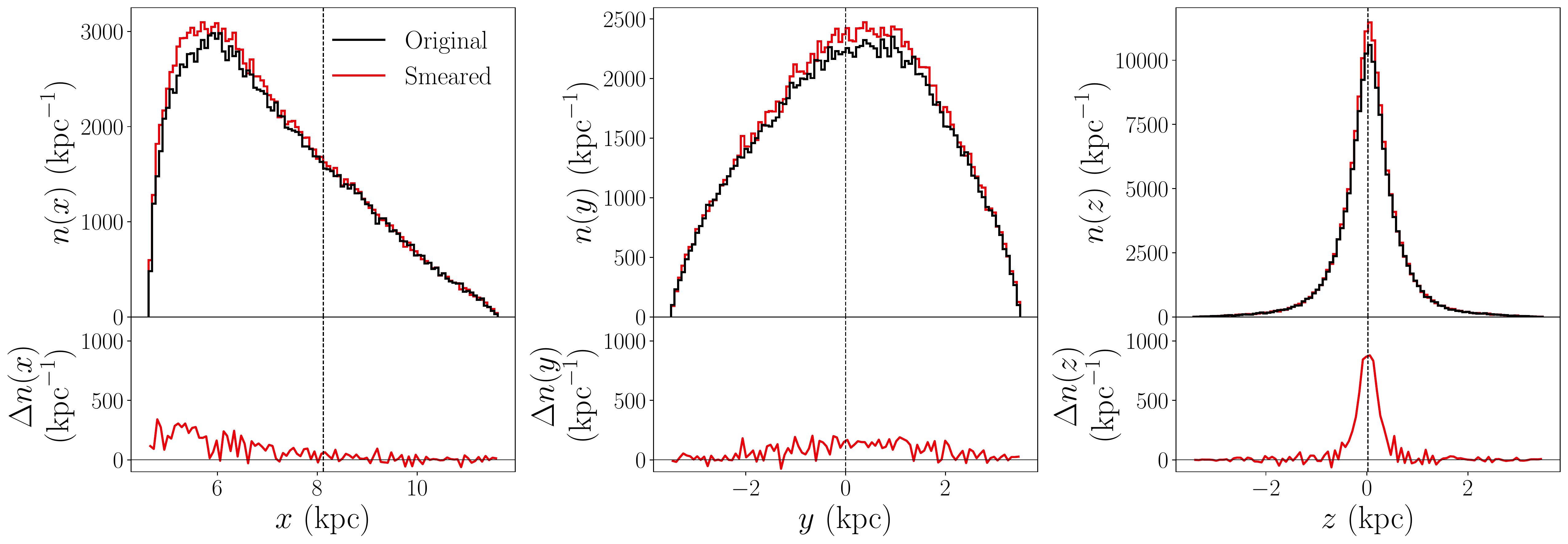}
\includegraphics[width=2.0\columnwidth]{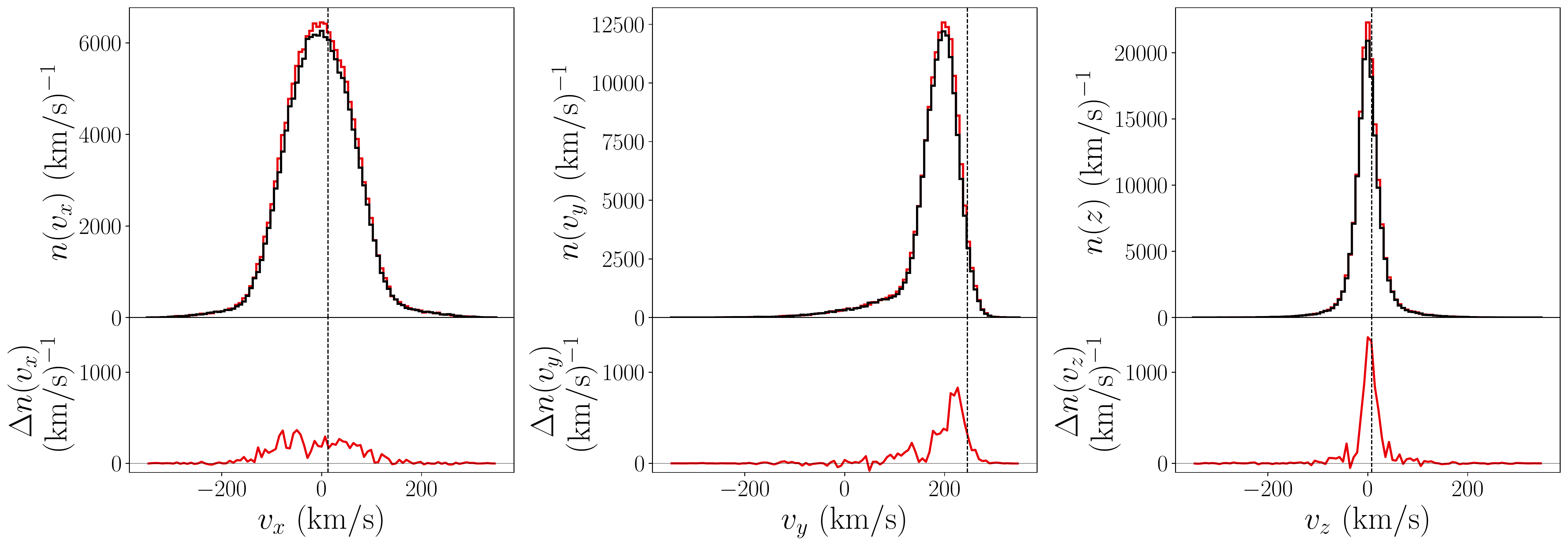}

\caption{One-dimensional number density distributions in galactocentric Cartesian coordinates of all stars within 3.5~kpc of the assumed Solar position $\vec{x}_\odot = (8.122,0.0,0.0208)$~kpc both before (``original,'' black lines) and after (``smeared,'' in red) measurement errors are applied. 
The subplots show the difference in density before and after the errors are applied: $\Delta f \equiv f_{\rm orig.} - f_{\rm smear.}$, with the red indicating the difference between the distribution without errors and the error-smeared distribution. The assumed Solar location and velocity are shown with vertical dashed lines. 
\label{fig:data_histograms}}
\end{figure*}

In Figure~\ref{fig:data_errors} we show the distribution of errors in the Sun-centered angular coordinates (plus distance and radial velocity), along with the resulting errors in the galactocentric Cartesian coordinates $(x,y,z,v_x,v_y,v_z)$. We show one-dimensional number density histograms along each Cartesian coordinate both before and after errors in Figure~\ref{fig:data_histograms}, and for the augmented dataset. As can be seen, even with 10\% errors in the measured distances, the truth-level (labeled ``original'' in our figures) and error-smeared data (labeled ``smeared'') are relatively similar. Note that the stellar density rises rapidly towards the Galactic Center and is highly peaked around the $z=0$; the drop off at low $x$ and large $|y|$ values seen in Figure~\ref{fig:data_errors} are the result of the projection of the spherical observational window onto the underlying distribution.

\section{Phase Space Density Through Flows \label{sec:flows}}

Recent advances in machine learning have made it possible to estimate phase space density in high-dimensional space. These techniques, when applied to stellar kinematics, allows us direct access to the phase space distribution of stars within the Milky Way. Though in this paper we will be concerned only with the phase space density as a function of Cartesian coordinates $\vec{x}$ and velocities $\vec{v}$, one can easily extend the machine learning frameworks to include other stellar properties of interest, such as metallicity. Of particular note is that the gradients of $f(\vec{x},\vec{v})$ can be accurately estimated even in datasets where the number of tracer stars is not large. In this work, we will demonstrate the use of these machine learned-estimates of phase space density in one particular application -- that of measuring acceleration and mass density though the Boltzmann Equation, which involves the derivatives of the phase space density $f(\vec{x},\vec{v})$ with respect to both $\vec{x}$ and $\vec{v}$. However, it is important to note that the fundamental quantity $f$ may have more versatile uses. 

To estimate the phase space density from data, we use normalizing flows, specifically Masked Autoregressive Flows (MAFs) \cite{papamakarios2018masked}. In this section, we describe normalizing flows and MAFs  (Section~\ref{subsec:flows}), our implementation and training on the data drawn from \textsc{h277} (Section~\ref{subsec:training}), our treatment of various sources of errors (Section~\ref{subsec:errors}), and finally comparisons of our trained density estimates with the data itself (Section~\ref{subsec:estimates}).

\subsection{Normalizing Flows \label{subsec:flows}}

A normalizing flow is a statistical model that represents a probability distribution depending on a set of variables (the input data) as a chain of simple differentiable and fully-bijective transformations of a known base distribution.
Let $\vec{h}$ be a random variable from the base distribution (which we take to be the unit normal distribution in this work), and $g$ is a bijection that maps $\vec{h}$ to our input variables $\vec{u}$ (in the context of the stellar tracer population, $\vec{u}$ will be the Cartesian positions and/or velocities of the stars).
Then, the probability density $p$ of $\vec{u}$ can be calculated as follows,
\begin{equation}
\vec{h} \rightarrow \vec{u} = g(\vec{h}), \quad 
p(\vec{u}) = p(g^{-1}(\vec{u})) \, \left| \det \frac{d g(\vec{h})}{ d\vec{h}} \right|^{-1}.
\end{equation}
Constructing a parametric model of bijection $g$ for arbitrary probability distribution is nontrivial, but a normalizing flow method achieves this by using a chain of simple\footnote{A ``simple'' bijection in the context of normalizing flows often means that its inverse and Jacobian determinant are easily computable.
} tractable nonlinear bijections $g_1, \cdots, g_N$ as the trasformation $g$,
\begin{equation}
g(\vec{h}) = g_N \circ \cdots \circ g_1(\vec{h}).
\end{equation}
The Jacobian of $g$ is simply a product of Jacobians of $g_1, \cdots, g_N$.
\begin{equation}
\det \frac{d g(\vec{h})}{ d\vec{h}} 
= 
\prod_{k=1}^N \det \frac{d g_{k}(\vec{h}_{k-1})}{ d \vec{h}_{k-1} }
\end{equation}
where $\vec{h}_k = g_k \circ \cdots \circ g_1(\vec{h})$, and $\vec{h}_0 = \vec{h}$.
A sufficiently long chain of transformations allows the normalizing flow to learn the transformation rules to map a complicated distributions in $\vec{u}$ from a simple (typically Gaussian) distribution in $\vec{h}$. For this reason, normalizing flows are considered an efficient model of general probability distributions.

In the case of modeling multivariate distributions, constructing simple multivariate bijections is not itself straightforward.
To overcome this, MAFs use autoregressive modeling of probability density, which utilizes the chain rule of probability in order to model a simple multivariate bijection as a product of simple conditioned univariate bijections.
The chain rule says that a joint probability $p(\vec{u})$ can be written in terms of conditionals $p(u_k | u_{1:{k-1}} )$,
\begin{equation}
p(\vec{u}) = p(u_1) \times p(u_2 | u_1) \times \cdots \times p(u_n | u_{1:{n-1}} ),
\end{equation}
where $u_{1:k}$ denotes a tuple $(u_1,\cdots, u_k)$ which is a subset of the elements of $\vec{u}$.
We only need to model conditioned univariate bijections for each $p(u_k | u_{1:{k-1}} )$ in order to construct a simple multivariate bijection.

In particular, we use affine MAFs whose conditional probability $p(u_k | u_{1:{k-1}})$ is Gaussian if the base distribution is the standard normal distribution.
The corresponding transformation from $\vec{h}$ to $\vec{u}$ is defined as follows.
\begin{equation}
h_k \rightarrow u_k = h_k \, \sigma(u_{1:k-1}) + \mu(u_{1:k-1}),
\end{equation}
where $\mu(u_{1:k-1})$ and $\sigma(u_{1:k-1})$ are the mean and standard deviation parameters of $p(u_k | u_{1:{k-1}})$,  respectively.

The parameter functions $\mu(u_{1:k-1})$ and $\sigma(u_{1:k-1})$ are modeled by a multilayer perceptron with masking, following the Masked Autoencoder for Density Estimation (MADE) \cite{pmlr-v37-germain15} approach.
The masking turns off connections to $k$-th output from $i$-th inputs with $i>k$, so that we may use this MADE block to build a model for the parameter functions of accumulated tuples $u_{1:k-1}$ as follows.
\begin{equation}
\mathrm{MADE}(\vec{u}) =
\left(\begin{matrix}
\mu & \log \sigma \\
\mu(u_1) & \log \sigma(u_1)  \\
\mu(u_{1:2}) & \log \sigma(u_{1:2})  \\
\vdots & \vdots \\
\mu(u_{1:n-1}) & \log \sigma(u_{1:n-1})  \\
\end{matrix}\right)
\end{equation}
We use $\log \sigma (u_{1:n-1})$ here to make $\sigma(u_{1:k-1})$ positive definite.

We emphasize several properties of these MAFs which are notable for our purposes. First, the fitting of the density estimator function to data is unsupervised; no assumption is placed on the values or functional form of the density. Rather than attempting to fit the data to some known distribution, the MAF minimizes a loss function
\begin{equation}
{\cal L} = - \sum_\alpha \log f(\vec{x}^\alpha,\vec{v}^\alpha),
\end{equation}
where the sum runs over the dataset $\{\vec{x}^\alpha,\vec{v}^\alpha\}$.\footnote{From a physics standpoint, it is perhaps interesting that this loss function is functionally the entropy of the dataset.}

Second, as the MAF output is a continuous function of the input coordinates, and the minimization of the loss function occurs through gradient-descent over the training, the trained neural network naturally tracks the derivatives of the density function with respect to the input dataset, allowing not just $f$ but $\vec{\nabla}_x f$ and $\vec{\nabla}_v f$ to be calculated. Finally, the MAF can be used as a sampler, generating pseudodata $(\vec{x},\vec{v})$ coordinates drawn from the underlying density distribution $f$.

One of our end goals is the calculation of the mass density $\rho$. This is related to the second derivatives of the phase space density via the Boltzmann and Poisson equations. To obtain continuous second derivatives, we use Gaussian Error Linear Unit (GELU) \cite{2016arXiv160608415H} for the activation functions of our MADE blocks. The GELU activation function is defined as
\begin{equation}
\mathrm{GELU}(x)= x \, G(x), \quad G(x) = \frac{1}{2} \left(1 + \mathrm{Erf} \frac{x}{\sqrt{2}} \right),
\end{equation}
where $G(x)$ is the cumuluative distribution function of the standard normal distribution.
GELU can be considered as one of smooth generalizations of Rectified Linear Unit (ReLU) \cite{Hinton_rectifiedlinear}
\begin{equation}
\mathrm{ReLU}(x) = x \max(x,0).
\end{equation}
The MADE block with GELU is then a smooth function so that its derivatives are well-defined.

\subsection{MAF Implementation and Training \label{subsec:training} }

Given the structure of the dataset we are interested in, where each star is characterized by a position $\vec{x}$ and velocity $\vec{v}$, and by the form of the Boltzmann Equation Eq.~\eqref{eq:boltzmann_mastereq}, in which the acceleration is a function of position only, we find it most convenient and accurate to write the total density distribution of the six-dimensional phase space in terms of the position-space density $\nu(\vec{x})$ and the conditional density $p(\vec{v}|\vec{x})$:
\begin{equation}
f(\vec{x},\vec{v}) = p(\vec{v}|\vec{x})\nu(\vec{x}). \label{eq:density_conditional}
\end{equation}
Note that the density $\nu$ is a number density of tracers, not the overall mass density $\rho(\vec{x})$ responsible for sourcing the gravitational potential (though of course in the simulated galaxy as in the Milky Way, the tracer stars contribute a non-negligible amount to the local mass density). The MAF can easily be modified to calculate $p(\vec{v}|\vec{x})$, by allowing the MADE blocks to take $\vec{x}$ as additional inputs whose connections to the probability density output are not masked.

We ensemble average our MAF results to reduce variance in our predictions.
We first train 10 MAFs using different random seeds for the parameter initialization, training/validation dataset split, and building mini-batches.
The two densities $\nu(\vec{x})$ and $p(\vec{v}|\vec{x})$ are then the averages of probabilities $\nu(\vec{x};s)$ and $p(\vec{v}|\vec{x};s)$ estimated by MAFs trained using a seed $s$:
\begin{eqnarray}
\nu(\vec{x}) 
& = &
\frac{1}{N}\sum_s \nu(\vec{x}; s), \label{eq:nu_ensembling}
\\
p(\vec{v} | \vec{x})
& = &
\frac{1}{N} \sum_s
p(\vec{v} | \vec{x}; s), \label{eq:p_ensembling}
\end{eqnarray}
where $N=10$ for this work.
Derived quantities such as the accelerations and mass densities are estimated using these averaged probabilities.

Our MAFs are built in \textsc{nflows} \cite{nflows} and \textsc{PyTorch} \cite{NEURIPS2019_9015}. Our architecture is described in Appendix~\ref{app:maf}. Prior to training, the position and velocity datasets are pre-processed so as to improve the ability of the MAF to fit the data. The preprocessing steps are described in Appendix~\ref{app:preprocess}.

We divide the total dataset into 80\% for training and 20\% for validation.
The training/validation dataset split is randomly assigned for each of the $N$ MAFs; as a result, our ensemble averaging also uses a Monte Carlo cross-validation. The network parameters are trained by minimizing the negative log-likelihood using the ADAM optimizer \cite{2014arXiv1412.6980K} with learning rate $10^{-3}$ and a minibatch size given by one-tenth of the training set. 
If the validation loss does not improve for 50 epochs, we repeat the same training algorithm with a reduced learning rate of $10^{-4}$ to fine-tune the network parameters. Training ceases once the validation loss does not improve for another 50 epochs at this reduced learning rate.

\subsection{Measurement Error Bias Correction}
\label{subsec:biascorr}

Mis-measurement of the position and velocity of the tracer stars will result in a bias in the MAF-derived phase space density $f(\vec{x},\vec{v})\equiv f(\vec{w})$ as the errors move stars from regions of high density into regions of low density. One may conceptualize this bias as a ``heating'' of the system due to measurement errors, which will increase the entropy of the system $S \equiv -\int d^6\vec{w} f \ln f$. While techniques such as extreme deconvolution \cite{Bovy_2011} exist to undo this measurement bias, as the errors in our {\it Gaia}-inspired simulated dataset are small near the Solar location we take a straightforward approach to the problem in this paper, leaving a more complete method for future work.

For our derivation, we assume that the errors in the phase space coordinates are small, that the error model changes slowly with position, and that we perfectly understand the error model. While only the last two of these assumptions are completely valid in the simulated dataset (and the last would not be perfectly true in the real data), these assumptions are approximately satisfied and allow for an analytic treatment in lieu of a more general deconvolutional approach which must be applied for large position-dependent errors.\footnote{Such deconvolutional methods would also assume that the modeling of the errors was perfectly understood.} Under these assumptions, generating a reperturbed dataset by randomly generating an additional error for each tracer star in the smeared dataset will, on average, shift the phase space density and its derivatives by the same amount as the shift between the original error-free dataset and the smeared data. By comparing the density and derivatives in the smeared and reperturbed datasets, this shift can be calculated and then subtracted. For more details, see Appendix~\ref{app:errors}.

\subsection{Error Propagation \label{subsec:errors} }

Like any data analysis, the sources of error in our machine-learned results must be quantified. We separate our sources of errors into two broad classes: statistical and systematic. Statistical errors arise from the specific initialization of our MAF architecture, and from the finite statistics of the dataset on which the MAFs are trained. Systematic errors originate from the measurement error in the smeared dataset.

The first source of statistical error is the variation of the MAF output due to variations in training caused by different initialization seeds. This variation can be estimated using the variation within the ensemble of 10 MAFs over which we have averaged in Eqs.~\eqref{eq:nu_ensembling} and \eqref{eq:p_ensembling}. 
The uncertainty on mean values of $f$, $\partial f/\partial x_i$, and $\partial f/\partial v_i$  are then the standard deviation of each quantities within the ensemble, divided by $\sqrt{N}$ (with $N=10$). We have verified that with $N=10$ MAFs, the initialization errors are always negligible compared to the other sources of error, so we will ignore them going forward. 

The second source of statistical uncertainty arises from the finite number of tracers in the dataset on which our MAFs are trained. To quantify this, we perform a bootstrap of the dataset. From the dataset containing $n$ tracer stars ($n=160,881$ in our smeared dataset), we draw $n$ tracers {\it with} replacement. We repeat this process to create 10 separate bootstrap datasets. Each of these datasets are then used as inputs to train new MAFs. The bootstrap error on a quantity then is the standard deviation of the mean of the 10 different values obtained from the 10 bootstrapped datasets. If the initial dataset was larger, then the bootstrapped datasets would similarly increase in size, and thus the 10 MAFs would converge on similar results; thus the bootstrap error is an estimate for the error in the results due to the finite statistics of the available data. 

Since any training of the bootstrap MAFs will themselves include the the variation of the ensemble averaging within the results from the bootstrap MAFs, rather than calculating these two forms of error separately, we can estimate a single ``statistical'' error by considering the variation in the results of 10 MAFs trained on 10 bootstrap datasets.

Next, we must consider systematic errors, which originate from the underlying measurement uncertainty. Although we have subtracted off the overall bias using the procedure described in Section~\ref{subsec:biascorr} and Appendix~\ref{app:errors}, there is still the variance that must be quantified.  In order to do so, we generate 10 new datasets from the error-smeared data -- drawing new random errors for each tracer star\footnote{It is at this point that our assumption that the error model is slowly varying with position comes in, as we are drawing a random error based on the perturbed position of the tracer not the original position.} -- and train new MAF networks on each reperturbed dataset. For each derived quantity, such as acceleration and mass density, we subtract the average difference between the results of the 10 reperturbed MAFs and the MAFs trained on the smeared dataset. The $1\sigma$ deviation of this subtraction we classify as a systematic error. See Appendix~\ref{app:errors} for more details. 

\subsection{MAF Density Estimates \label{subsec:estimates} }

\begin{figure*}
\includegraphics[width=2.0\columnwidth]{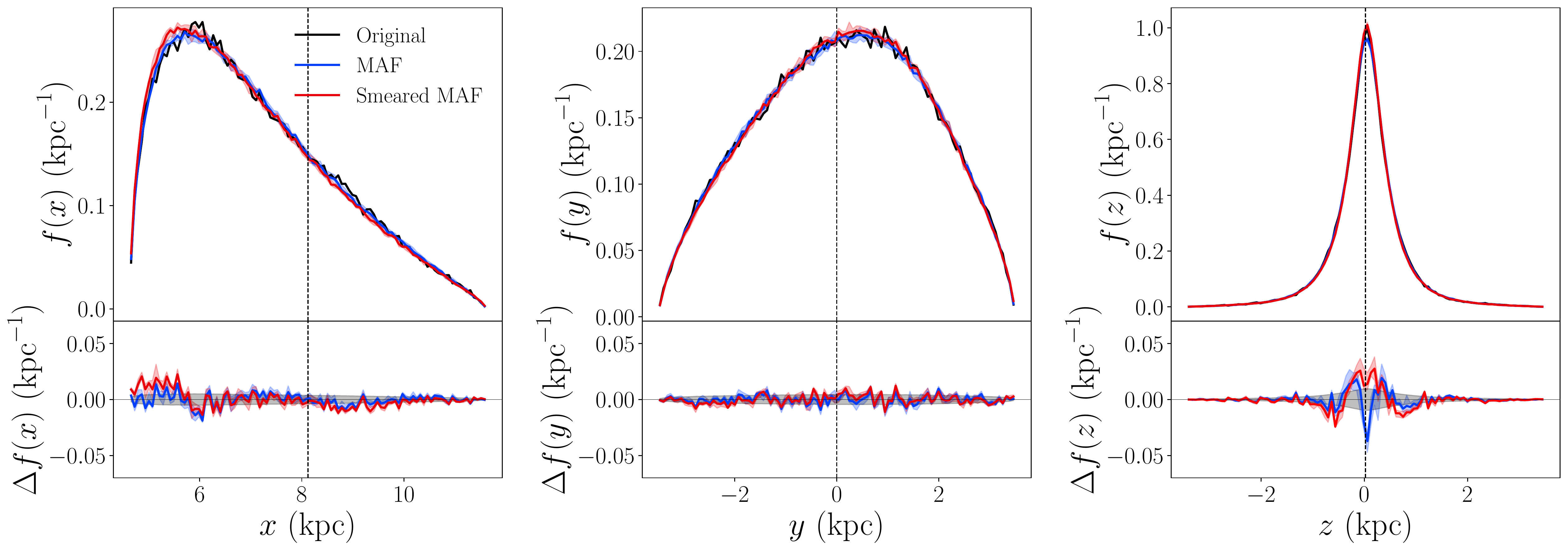}
\includegraphics[width=2.0\columnwidth]{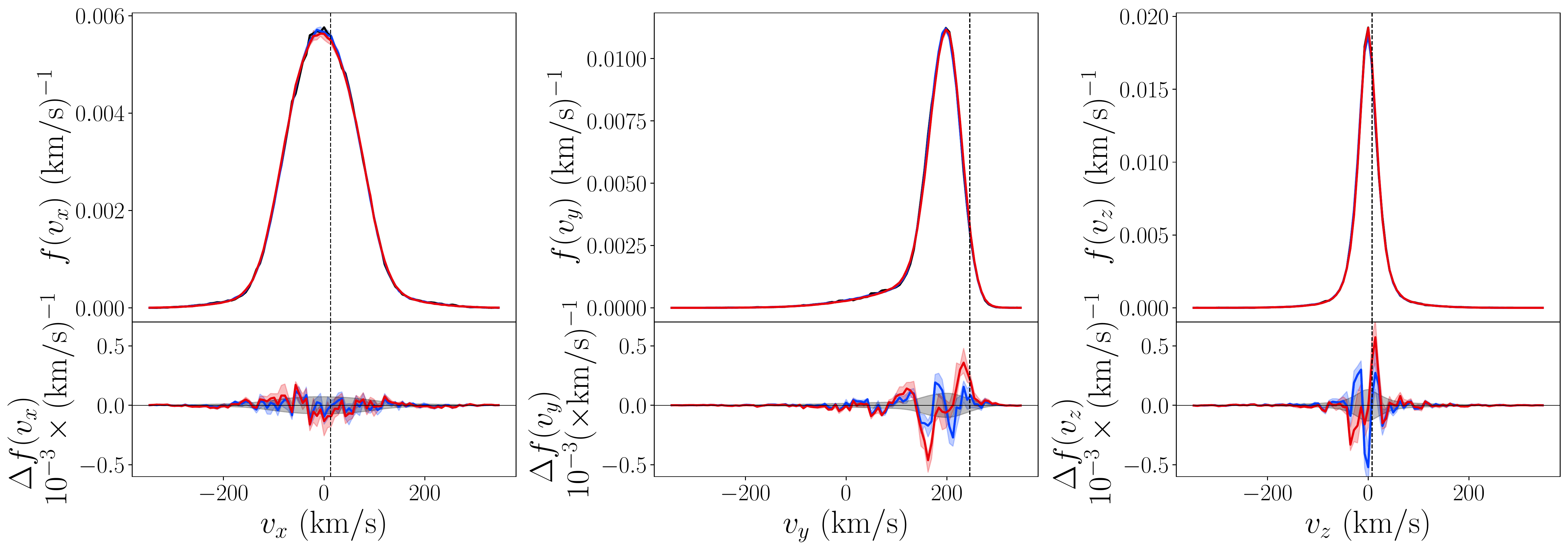}

\caption{One-dimensional density distributions in galactocentric Cartesian coordinates of $10^6$ stars sampled from 10 MAFs trained on datasets of stars within within 3.5~kpc of the Solar position $\vec{x}_\odot = (8.122,0.0,0.0208)$~kpc. Sampled stars from MAFs trained on the original dataset before errors are applied are shown in blue, and after error smearing in red, while the original (unsmeared) dataset is shown in black. Shaded regions indicate the $1\sigma$ error from the ensemble of 10 MAFs. Vertical dashed lines show the Solar velocity. 
The subplots show the difference in in the one-dimensional histogram between the sampled stars and the original distribution of the data without errors:  $\Delta f \equiv f_{\rm MAF} - f_{\rm true}$ (where $f_{\rm true}$ is obtained via one-dimensional binning). The grey ellipse corresponds to expected $1\sigma$ Gaussian errors based on the number of original dataset stars in each bin, and the colored shading indicates $1\sigma$ errors from the ensemble of 10 MAFs. \label{fig:density_output}
}
\end{figure*}

To demonstrate the accuracy of the MAFs in reconstructing the phase space density of the original datasets, as well as the effect of each of the types of errors we consider, we sample each pair of position and velocity MAFs to generate pseudodata ``stars," drawn from the MAF-learned density. In Figure~\ref{fig:density_output} we show one-dimensional histograms of each kinematic phase-space variable drawn from an ensemble of MAFs (the equivalent of Figure~\ref{fig:data_histograms}). By eye, these histograms are nearly indistinguishable from the datasets themselves, and so we also show in Figure~\ref{fig:density_output} the difference between the datasets and the MAF reconstructions, along with the Gaussian error envelope in each bin based on the number of stars in the original dataset.

\begin{figure*}
\includegraphics[width=2.0\columnwidth]{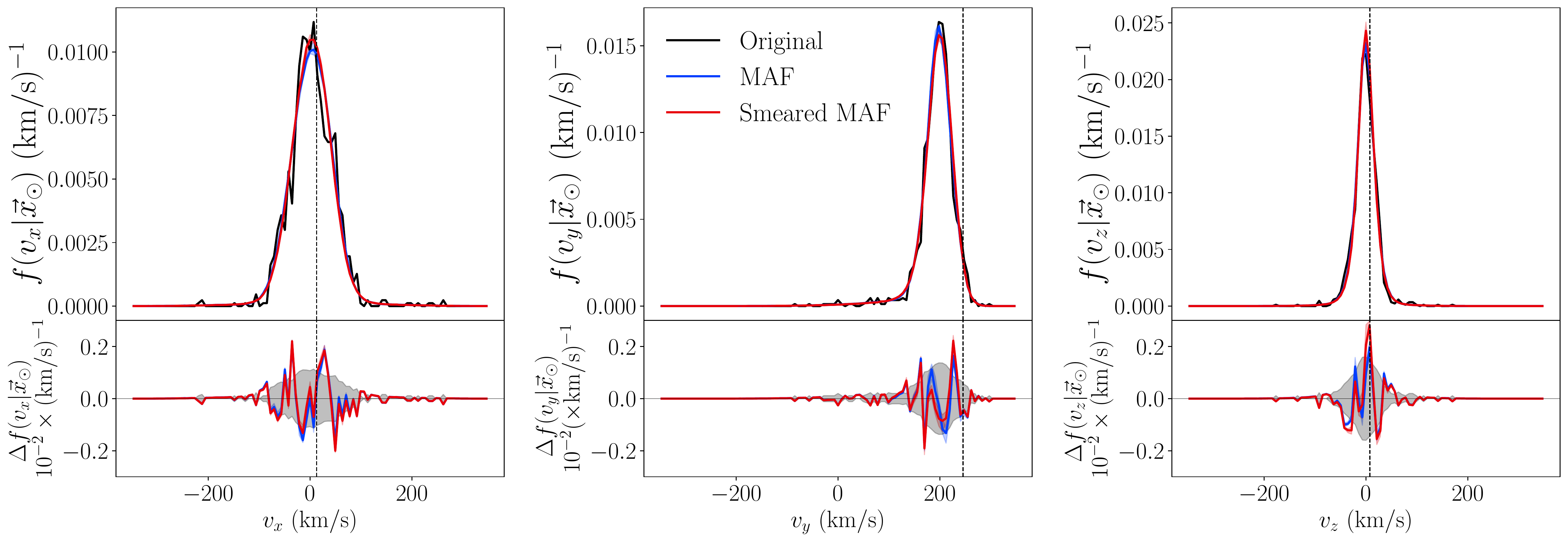}

\caption{One-dimensional density distributions with respect to Cartesian velocities in galactocentric coordinates of $10^6$ stars sampled from 10 velocity MAFs 
conditioned on the Solar position $\vec{x}_\odot = (8.122,0.0,0.0208)$~kpc. 
Sampled stars from MAFs trained on the original dataset before errors are applied are shown in blue and after error smearing in red. 
The subplots show the difference in in the one-dimensional histogram between the sampled stars and the distribution of the original data without errors, taking stars within $0.5$~kpc of the Solar position:  $\Delta f \equiv f_{\rm MAF} - f_{\rm true}$ (where $f_{\rm true}$ is obtained via one-dimensional binning of all stars within 0.5~kpc of the Solar location). The grey ellipse corresponds to $1\sigma$ Gaussian errors based on the number of dataset stars in each bin, and the colored shading indicates $1\sigma$ errors from the ensemble of 10 MAFs \label{fig:velocity_density_output}  }
\end{figure*}

These one-dimensional histograms indicate that, when taking the entire range of the dataset, the MAFs are largely successful in learning the dataset density. However, the approach to solving the Boltzmann Equation we will take in Section~\ref{sec:boltzmann} requires accurate measurements of the velocity distribution conditioned on the position. In Figure~\ref{fig:velocity_density_output}, we show the velocity distributions of pseudodata sampled from the MAFs conditioned on a specific location: the Sun's location $\vec{x}_\odot = (8.122,0,0.0208)$~kpc.

We then compare to the velocity distributions of the dataset (selecting all stars within $0.5$~kpc of the Solar location). As can be seen, the errors for velocity conditioned on location are an order of magnitude larger than those for the entire dataset, reflecting the fact that the fewer tracers are available to constrain the velocity distribution when conditioning on position, versus sampling the velocity distribution over the entire observation window.

\section{Acceleration From the Boltzmann Equation \label{sec:boltzmann}}

Having introduced the simulated dataset and our method of learning the density and derivatives of that dataset, we can now turn to determining the acceleration and overall density in which these tracer stars were evolving.  Our initial goal is to determine the potential gradient at any position $\vec{x}$ which lies within the domain of our tracer dataset, starting with the Boltzmann Equation in Cartesian coordinates Eq.~\eqref{eq:boltzmann_mastereq}.  As pointed out in Ref.~\cite{2020arXiv201104673G,2021arXiv211207657N,2021MNRAS.506.5721A}, as $\Phi$ is a function of $\vec{x}$ only, multiple velocities sampled from $p(\vec{v}|\vec{x})$ at the same $\vec{x}$ must obey the Boltzmann Equation given the same acceleration $\vec{a} \equiv -\partial\Phi/ \partial \vec{x}$. 

As the MAFs do not perfectly reconstruct the phase space densities and local equilibrium is not perfectly achieved, we expect the time derivative of the phase-space density $df/dt$ to differ from zero, star-by-star. However, under the assumptions of approximate equilibrium and that the MAFs are approximately correct, this time derivative will be on average zero. We can then obtain an estimator for the the acceleration at a location $\vec{x}$ by minimizing the mean squared error (MSE) for $df/dt$:
\begin{eqnarray}
\mathcal{L}_{\mathrm{MSE}}(\vec{x}) 
 &= &
\frac{1}{N} \sum_{\beta=1}^N \; 
\left|
\frac{d f}{ d t}
\right|^2_{\vec{v} = \vec{v}^{\beta}} \nonumber \\
 &= &
\frac{1}{N} \sum_{\beta=1}^N \; 
\left|
v_i \frac{\partial f}{\partial x_i} + a_i \frac{\partial f}{\partial v_i}
\right|^2_{\vec{v} = \vec{v}^{\beta}} \label{eq:boltzmann_dln}
\end{eqnarray}
where the sum runs over $N$ sampled star velocities at $\vec{x}$, and we have replaced $\partial\Phi/\partial\vec{x}$ with $-\vec{a}$. 

Taking a derivative of the MSE loss function with respect to acceleration allows us to determine the minimum analytically:
\begin{equation}
\frac{N}{2} \frac{ \partial \mathcal{L}_{\mathrm{MSE}}}{\partial a_i}
=
\sum_{\beta=1}^N
\left[
\left(
v_j \frac{\partial f}{\partial x_j} + a_j \frac{\partial f}{\partial v_j}
\right) \frac{\partial f}{\partial v_i}
\right]_{\vec{v} = \vec{v}^{\beta}}
\end{equation}
This is a matrix equation of the form
\[
V_i +M_{ij}a_j =0
\]
where
\begin{eqnarray}
M_{ij} 
& = & 
\sum_{\beta=1}^N 
\left[
\frac{\partial f}{\partial v_i} \frac{\partial f}{\partial v_j}
\right]_{\vec{v} = \vec{v}^{\beta}}, \label{eq:M_def}
\\
V_i 
& = &
\sum_{\beta=1}^N 
\left[
\left( v_j \frac{\partial f}{\partial x_j} \right)
\frac{\partial f}{\partial v_i}
\right]_{\vec{v} = \vec{v}^{\beta}}, \label{eq:V_def}
\end{eqnarray}
which can be analytically solved by inverting $M_{ij}$. 
This mean-squared-error solution can also be motivated by treating $df / dt = 0$ as a matrix equation in mixed coordinate- and sampled velocity-space, and solving for $\vec{a}$ using the inverse of the rectangular matrix $K_{i\beta} = \left. \partial f/ \partial v_i \right|_{\vec{v}=\vec{v}^{\beta}}$.

The sampled velocities that enter into the sums of Eqs.~\eqref{eq:M_def} and \eqref{eq:V_def} are drawn from the MAF-derived $p(\vec{v}|\vec{x})$. One advantage of this sampling is that the samples at the tails of the $p(v|x)$ distribution contribute less to the MSE loss, minimizing the effects of the often-poor fits in the tails. That said, the MAF will occasionally generate high-velocity samples which have no nearby supporting training samples in the phase space. This occurs because the MAF covers the full Euclidean space of velocities, yet the data to which the MAF was fit does not have any constraints on large generated velocities. These high-velocity samples typically result in divergent position and velocity derivatives of $p(\vec{v}|\vec{x})$. To avoid these spurious contributions to the acceleration solution, we discard sampled $\vec{v}^\beta$ which have magnitudes greater than $0.8\times v_{\rm max}$, where $v_{\rm max}$ is the maximum speed measured for any star in the training dataset.

\begin{table}
\caption{
Simulation-truth and estimated accelerations at the Solar location. 
We also show the uncertainties discussed in Section~\ref{subsec:errors}. 
For the simulation-truth, the systematic uncertainly corresponds to the averaging of the simulation-truth acceleration over a Gaussian of radius 0.2~kpc, comparable to the resolution limit of the simulation.
Note that, due to the assumption of axisymmetry, the acceleration component $a_y$ exactly vanishes in the Jeans analysis.
\label{tab:acc_at_sun}
}
\begin{center}
\begin{ruledtabular}
\begin{tabular}{llccc}
\multirow{2}{*}{component} & \multirow{2}{*}{dataset} & \multicolumn{3}{c}{acceleration $\rm (kpc / Gyr^2)$} \\
& & & (stat.) & (syst.) 
\\
\hline
\multirow{4}{*}{$a_x$} 
& sim.-truth & $-5608$ & -- & $\pm 193$ \\
& original & $-5672$ & $\pm \phantom{0}43$ & -- \\
& smeared & $-5597$ & $\pm \phantom{0}67$ & $\pm \phantom{0}40$\\
& Jeans & $-5305$ & $\pm \phantom{0}67$ & $\pm \phantom{0}19$ \\
\hline
\multirow{4}{*}{$a_y$}
& sim.-truth & $\phantom{-00}41$ & -- & $\pm 116$ \\
& original & $\phantom{-0}344$ & $ \pm \phantom{0}51$ &-- \\
& smeared & $\phantom{-0}237$ & $\pm \phantom{0}45$ & $\pm \phantom{0}29$\\
& Jeans & $\phantom{-000}0$ & -- &  --\\
\hline
\multirow{4}{*}{$a_z$}
& sim.-truth & $\phantom{-00}86$ & -- & $\pm 381$ \\
& original & $\phantom{-00}60$ & $\pm \phantom{0}30$ & -- \\
& smeared & $\phantom{-00}72$ & $\pm \phantom{0}30$ & $\pm \phantom{0}13$\\
& Jeans & \phantom{00}$-38$ & $\pm \phantom{00}3$  & $\pm \phantom{00}1$ \\
\end{tabular}
\end{ruledtabular}
\end{center}
\end{table}

We compare our MAF-derived accelerations with the simulation-truth accelerations. The latter is calculated via explicit sum using \textsc{PynBody} routines over the gravitational accelerations from all particles in the $N$-body simulation (with a softening length set by the individual particle resolutions). 
At the Solar location in the center of our spherical dataset, the acceleration in the simulation is shown in Table~\ref{tab:acc_at_sun}.
Since our galaxy simulation has a finite resolution, we report the mean and standard deviations of accelerations evaluated over the positions perturbed by a Gaussian whose mean and standard deviation are the location of the Sun and the softening length,\footnote{We use 0.2~kpc for the standard deviation, compared to the median softening length for particles in the simulation of 0.173~kpc.} respectively. The narrowness of the disk compared to the resolution length of the simulation results in a comparatively large error on the averaged  simulation-truth vertical acceleration component $a_z$.
This standard deviation of acceleration can be regarded as a systematic uncertainty of the simulation-truth accelerations.
From Table~\ref{tab:acc_at_sun}, we see that the MAF results largely agree within errors with the simulation truth.

\begin{figure*}
\includegraphics[height=0.3\textheight]{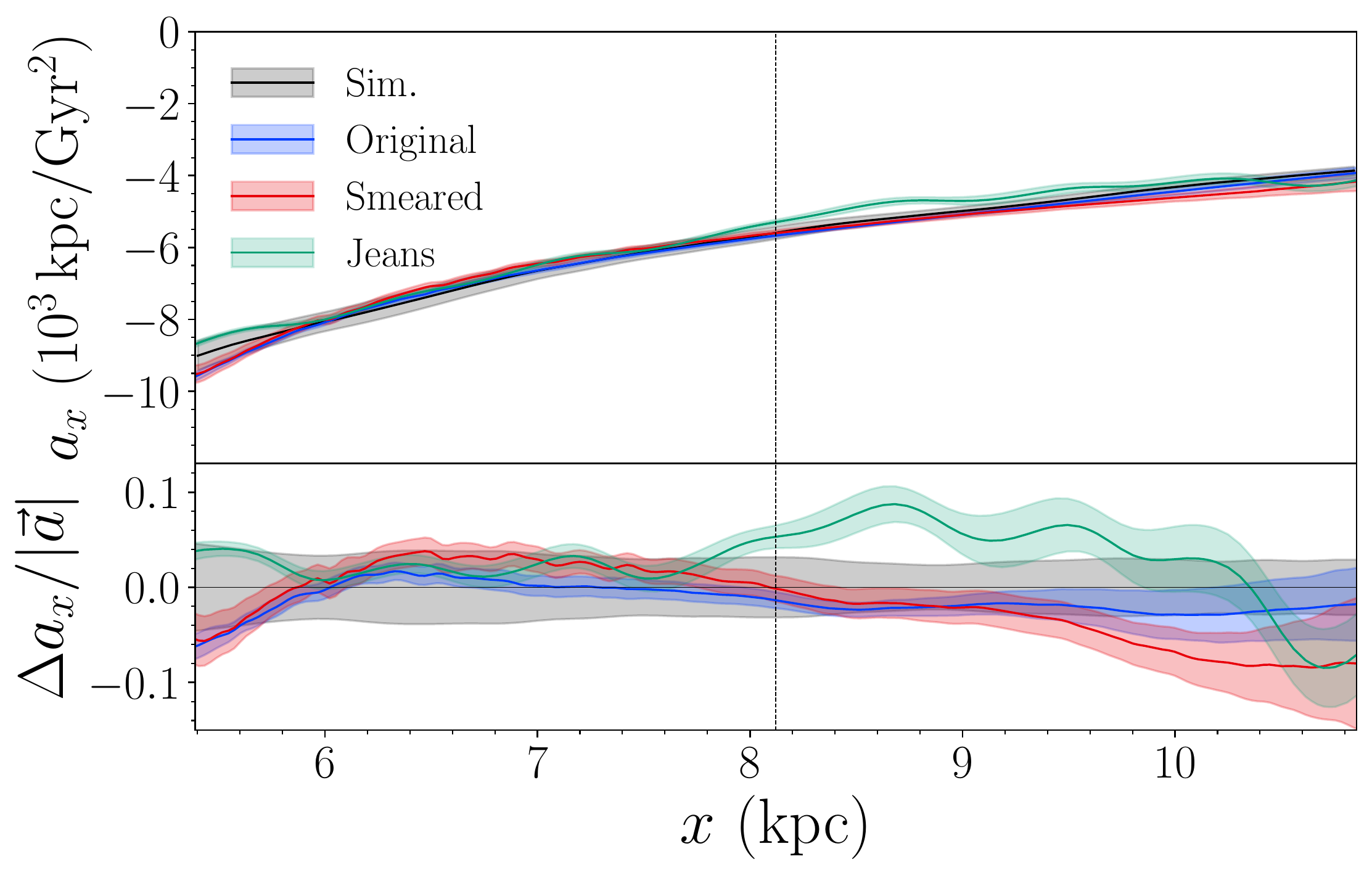}\\
\includegraphics[height=0.3\textheight]{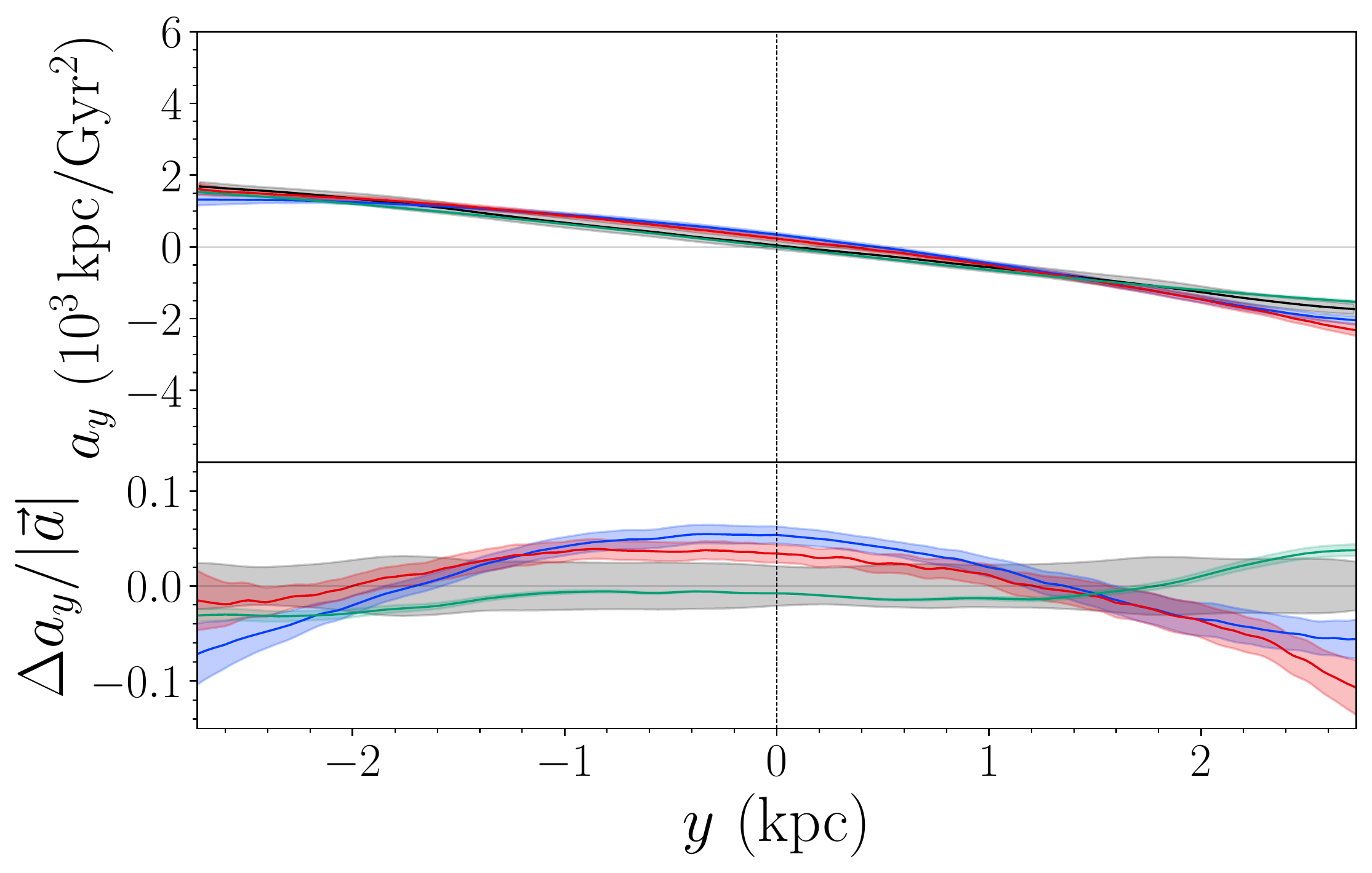}\\
\includegraphics[height=0.3\textheight]{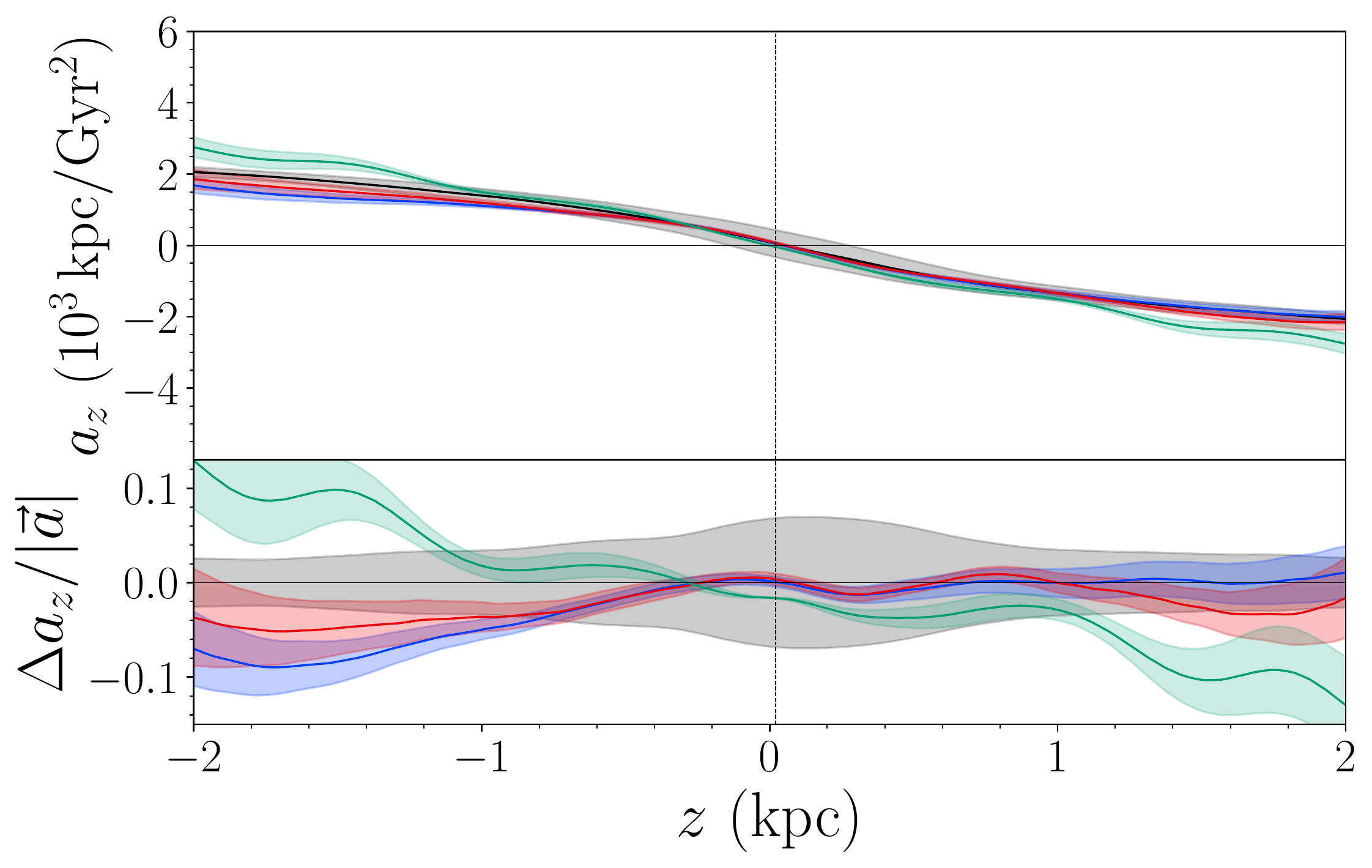}
\caption{MAF accelerations $a_x$ as a function of location along the $x$-axis with $y=z=0$ (top), $a_y$ as a function of location along the $y$-axis with $x=8.122$~kpc and $z=0$ (middle), and $a_z$ as a function location along the $z$-axis  with $x=8.122$~kpc and $y=0$ (bottom) for the original dataset (blue) and the average over error-smeared dataset (red), as compared to the simulation-truth (black). The accelerations derived using the Jeans analysis approach are shown in green. Fractional deviations from the true acceleration $\Delta a_i/|\vec{a}|$ are shown below each main plot. 
The shaded regions indicate the total $1\sigma$ error estimates. \label{fig:acceleration_1d}}
\end{figure*}

To investigate whether this agreement is possible for points other than the Solar location at the center of our sphere of simulated stars, in Figure~\ref{fig:acceleration_1d} we show $a_{x,y,z}$ vs $x,y,z$ respectively (passing through the Solar location), as calculated by the MAF for both the original (error-free) and error-smeared datasets. We show for comparison the simulation-truth accelerations. Though all three acceleration vector components at every location are relevant for the calculation of mass density (as we will describe in Section~\ref{sec:density}), in the main text we show only these selected components for clarity. The plot of all three acceleration components along the $x$, $y$, and $z$ axes can be found in Appendix~\ref{app:accelerations}.

As can be seen, we find agreement at the $\lesssim 10\%$ level for all components of the accelerations along the three axes. For much of the dataset, we find agreement to within $2\sigma$ of the MAF errors. The major exceptions 
are at high $x$-values and large $|z|$ values (far from the disk). These high $|z|$ and large $x$ regions are the least-well populated by tracer stars, and so these poor fits reflect low statistics. 

We can compare our MAF-derived results to the accelerations obtained via a Jeans equation analysis that does not use deep learning techniques. 
The details of our Jeans analysis approach are described in Appendix~\ref{app:jeans}. Note that the Jeans analysis assumes perfect axisymmetry, and so has very small accelerations (with accompanying small error bars) for $a_y$ along the $y$-axis. 

\subsection{Local Departure from Equilibrium \label{subsec:equilibrium}}

In the next section of this paper, we calculate the mass density of the galaxy from the accelerations derived from the phase space density of the tracer stars. However, we first consider what additional information about our simulated galaxy can be extracted from the phase space density assuming the accelerations are known from a separate source. In particular, we use the truth-level accelerations from the simulation to calculate the deviation from equilibrium $df/dt$ using the Boltzmann Equation Eq.~\eqref{eq:boltzmann_mastereq}. As we are interested primarily in true deviations from equilibrium, for this subsection we only consider the original dataset without measurement errors added.

We set up a grid of positions on the galactic plane ($z=0$) within our observational window of 3.5~kpc around the Solar location. For each position point, we sample 1000 velocities drawn from a uniform distribution over a spherical region in velocity-space with $|v| < 0.8\times v_{\rm max}$ (where $v_{\rm max}$ is the maximum velocity found in the dataset), and calculate $\partial f/\partial x_i$ and $\partial f/\partial v_i$ for each position and velocity averaged over the 10 MAFs. From this set of sampled velocities, we then calculate the time derivative of the position phase-space density:
\begin{equation}
\begin{multlined}
\frac{d\nu (\vec{x})}{dt} \equiv \int d^3 v \frac{df(\vec{x},\vec{v})}{dt} = \\
 -\frac{1}{N} \sum_{\alpha=1}^N \left[ v_i^\alpha \frac{\partial f(\vec{x},\vec{v}^{\,\alpha})}{\partial x_i}+ a_{i,{\rm sim}}(\vec{x}) \frac{\partial f(\vec{x},\vec{v}^{\,\alpha})}{\partial v_i} \right].
\end{multlined}
\end{equation}
Here the sum runs over the velocities sampled from the uniform distribution.

\begin{figure*}
\includegraphics[width=2.0\columnwidth]{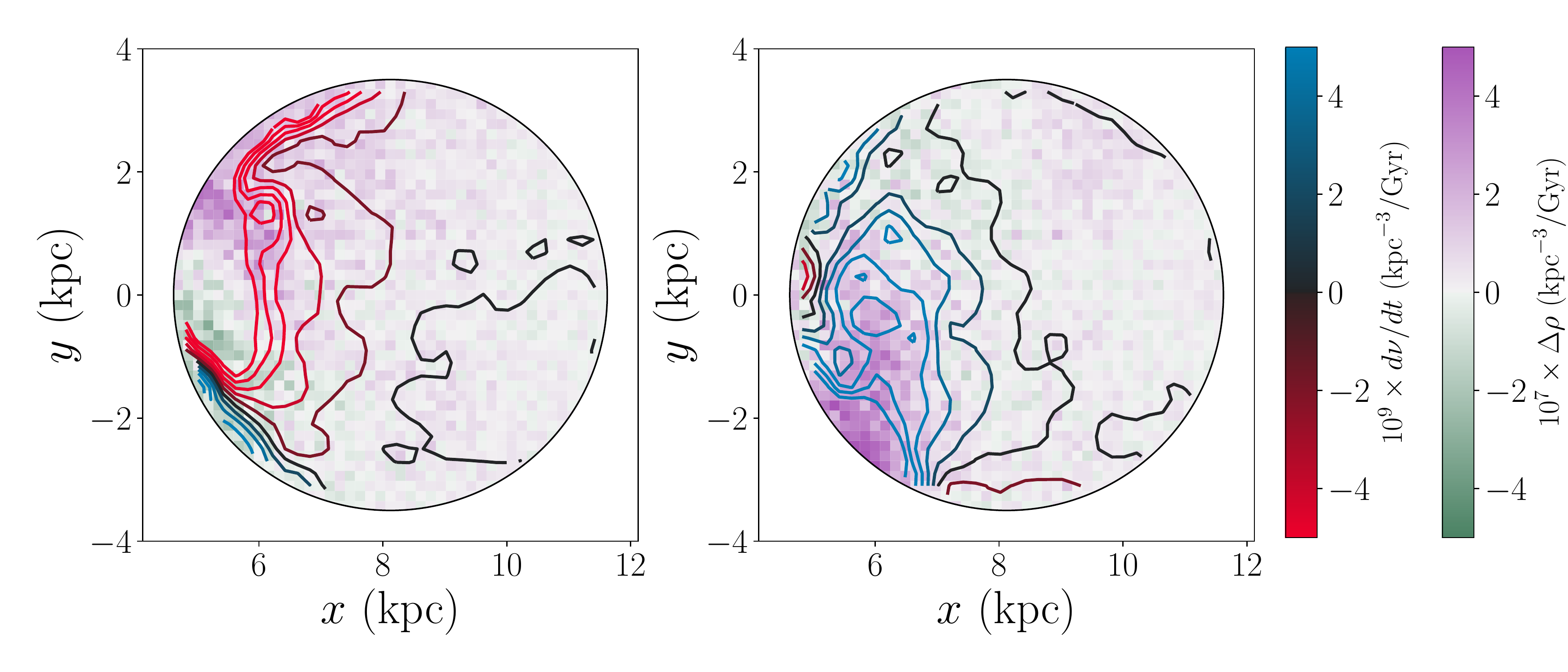}

\caption{Contour plot of $d\nu/dt$ in the galactic disk (blue-to-red contours) overlaid a histogram of the deviation from the average mass density of stars $\Delta \rho_{\rm stars}(\vec{x})$ for stars with $|z|<0.5$~kpc (green-to-purple density plot). The region centered on the  Solar location assumed throughout the rest of this paper is shown in the left. The right panel shows an alternate Solar location (rotated $55^\circ$ around the galactic center from the left plot) which contains more of a spiral arm. \label{fig:equilibrium} 
}
\end{figure*}

In Figure~\ref{fig:equilibrium}, we show maps of $d\nu/dt$ for two choices of Solar location. In the left panel, we show our canonical choice of the Solar location (our standard choice used throughout the rest of the paper). The right panel is centered on an alternate choice for the Solar location ($55^\circ$ around the galactic disk from the canonical center, at the same radius from the galactic center). Overlaid on these $d\nu/dt$ contours is the stellar mass density, subtracting off the stellar mass density at each radius, averaged over the galaxy: 
\begin{equation}
\Delta \rho_{\rm stars}(\vec{x}) \equiv \rho_{\rm stars}(\vec{x})-\bar{\rho}_{\rm stars}(r),
\end{equation}
where the mass density $\rho(\vec{x})$ is obtained through the binning of the simulation-truth star particles with  $|z| <0.5$~kpc, and the averaged density $\bar{\rho}(\vec{x})$ is obtained from binning over all stars in the disk at a given radius $r$. This density difference highlights the presence of the spiral arms, which appear as upward fluctuations.

While our canonical Solar location contains the edge of a spiral arm in the upper left of the 3.5~kpc observation window (visible as the purple region in the left panel of Figure~\ref{fig:equilibrium}), the alternate location contains more of an arm (seen in the lower left corner of the right-hand panel). As can be seen, in both cases, the $d\nu/dt$ calculated using the MAF densities and the truth-level accelerations trace the local departure from equilibrium in the disk stars which mark the passage of the spiral arms.

In the real Milky Way galaxy, many of the currently available acceleration measurements assume equilibrium, but there are methods -- such as pulsar and binary timing \cite{2021ApJ...907L..26C,2022ApJ...928L..17C} -- which provide probes of acceleration independent of the Boltzmann Equation. The accuracy and coverage of these techniques will only improve with time. With such information, the access to the phase space density provided by flow architectures could provide direct measures of the local equilibrium of the Milky Way's stellar population, as shown here. Alternatively, if one assumes a constant rotation of the spiral arms and no other time-dependent effects, the local change in the phase-space density of the disk stars could be derived from the MAF results and incorporated into the Boltzmann Equation. We will consider this possibility in a later work.  Finally, as discussed in Section~\ref{sec:data}, these out-of-equilibrium effects in the tracer star population could be mitigated if we considered only halo stars, rather than the majority disk star tracer population used here. This is not possible given the simulation at hand, as there are too few halo star particles within the observation window to allow for accurate training of the MAF.

\section{Mass Density from Accelerations  \label{sec:density}}

With an acceleration calculation in hand, we continue our derivation of the total mass density. 
To obtain the mass density from our numeric solutions of accelerations $\partial\Phi/\partial\vec{x}$ through the Poisson Equation Eq.~\eqref{eq:poisson}, we must take a further derivative with respect to position and compare it with the ``true'' mass density from simulation. We obtain mass densities from the MAF-derived accelerations by convolving the accelerations over a position-space kernel function, which avoids an explicit finite-difference calculation. 

The divergence of the acceleration convolved over a kernel function $K(\vec{x})$ and integrated over some volume can be written as 
\begin{equation}
\begin{multlined}
 \int d^3\vec{x}' (\vec{\nabla}\cdot \vec{a})(\vec{x}') K(\vec{x}-\vec{x}') = \oint d^2\vec{x}' (\hat{n}\cdot\vec{a})(\vec{x}') K(\vec{x}-\vec{x}') \\
-\int d^3\vec{x}' \vec{a}(\vec{x}') \cdot \vec{\nabla}K(\vec{x}-\vec{x}'). \label{eq:acceleration_integration}
\end{multlined}
\end{equation}
Here, the $\oint$ integral is over the surface of the volume of integration, with the unit vector $\hat{n}$ pointing perpendicular to the surface. 
Importantly, in this way we do not have to take finite differences of $\vec{a}$, which can introduce significant numeric instabilities from outliers.
We choose as our kernel $K$ a Gaussian with standard deviation $\vec{\sigma}$, truncated at two standard deviations. As the truncated Gaussian kernel does not go to zero on the boundary of an arbitrary volume, the boundary term is typically non-zero and must be included in the calculation.

Eq.~\eqref{eq:acceleration_integration} can be evaluated over the truncated Gaussian kernel using Monte Carlo sampling within the volume and on the surface: 
\begin{eqnarray}
4\pi G \rho(\vec{x}) 
& = & 
\frac{S}{N} \sum_{\alpha=1}^N (\hat{n} \cdot\vec{a})(\vec{x}^\alpha)K(\vec{x}^\alpha) \nonumber 
\\
&  & \;\; 
-\frac{1}{N} \sum_{\beta=1}^N \vec{a}(\vec{x}^\beta) \cdot \frac{(\vec{x}-\vec{x}^\beta)}{\vec{\sigma}^2},
\end{eqnarray}
where $S$ is the surface area of the volume centered on $\vec{x}$. The sampled points for the surface integral $\{\vec{x}^\alpha\}_{\alpha=1}^N$ are drawn from uniform distribution on the surface.
The dataset for the volume integral,  $\{\vec{x}^\beta\}_{\beta=1}^N$, is drawn from the Gaussian kernel $K(\vec{x} - \vec{x}')$ centered at $\vec{x}$.
With a finite number of samples, the acceleration components that pierce the surface but pass through and contribute zero net flux may not perfectly cancel due to small statistics. We avoid this error by explicitly symmetrizing the sum over the center location $\vec{x}$.

When selecting the size $\vec{\sigma}$ of the truncated Gaussian kernel, we must consider the resolution length of the \textsc{h277} simulation.
Though \textsc{PynBody}  reports mass density at individual arbitrary positions, this result is inherently noisy over scales near the resolution limit, as the mass density which sources gravity within the $N$-body simulation is a weighted sum of delta functions smoothed over this finite resolution.
To facilitate direct comparison between the simulation-truth mass density and the results of the MAF calculations, we choose to convolve the mass density of \textsc{h277} with the same truncated Gaussian kernel as we use for the finite difference calculation. 
A Gaussian standard deviation larger than the resolution limit of $0.2$~kpc would be ideal; however the change in density in the $z$ direction of the simulation (perpendicular to the disk) occurs over length scales equivalent to the resolution. 
We therefore choose to integrate over an anisotropic Gaussian, with $\sigma_x = \sigma_y = 1$~kpc, and $\sigma_z = 0.2$~kpc. 
We sample 6400 positions, which corresponds to $N=3200$, for the Monte Carlo integration, to achieve 2.5\% precision. 
For error estimations, we sample 1600 positions to achieve 5\% precision.

\begin{figure*}
\includegraphics[height=0.3\textheight]{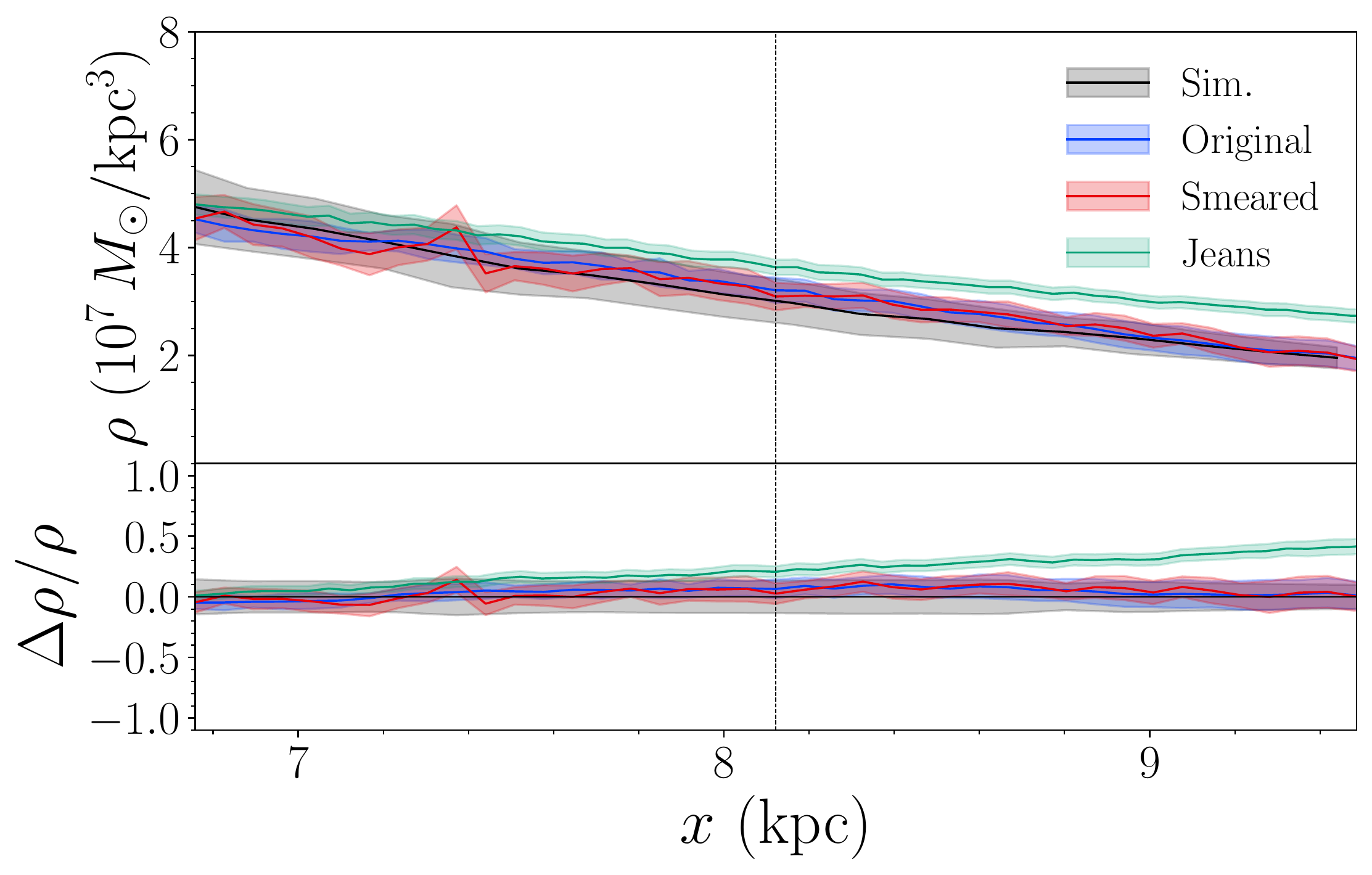}\\
\includegraphics[height=0.3\textheight]{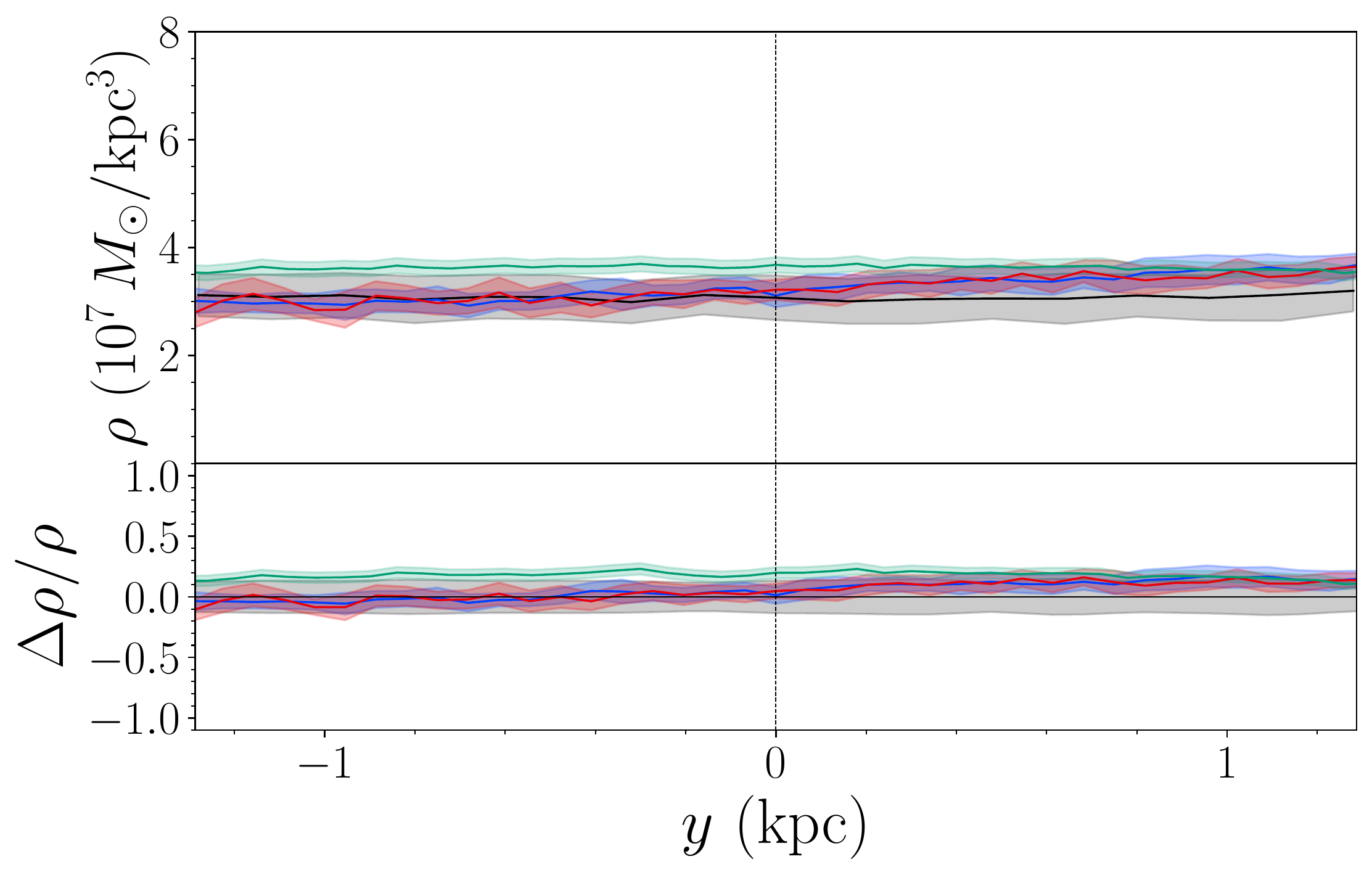}\\
\includegraphics[height=0.3\textheight]{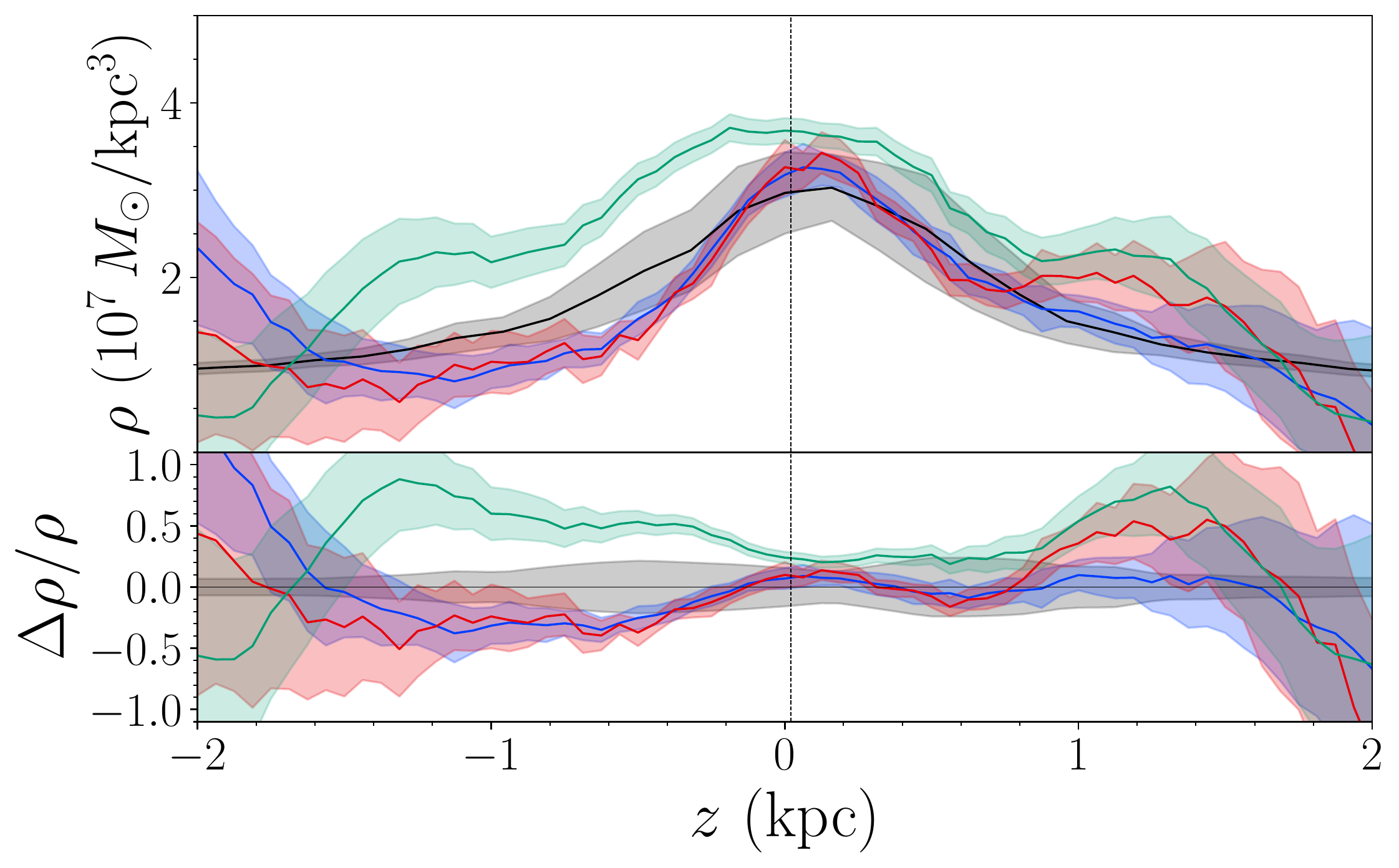}
\caption{MAF mass densities learned from the original dataset (blue) and averaged error-smeared dataset (red), as compared to the simulation-truth results (black), calculated for points along the $x$-axis with $y=z=0$ (top panel), along the $y$-axis through $8.112$~kpc and $z=0$ (middle panel), and along the $z$-axis with $x=8.122$~kpc and $y=0$ (bottom panel) using an anisotropic Gaussian kernel (see text). The density calculated using the Jeans method is shown in green. Fractional deviations from the central simulation-truth value are shown below each main plot. Shaded red and blue regions indicate the combined $1\sigma$ error. Black shaded regions show $1\sigma$ spread of simulation mass densities over the sampled kernel. \label{fig:density_1d} } 
\end{figure*}

In Figure~\ref{fig:density_1d}, we show the mass densities calculated from the MAF, trained on either the original or smeared dataset and using the anisotropic kernel, as compared to the simulation-truth values. As can be seen, inside the galactic disk, the MAF results agree with the true mass density within errors (and variations of the true densities within the sampling kernel). At large galactic radii, the smeared dataset values deviate somewhat from the true values. Both the smeared and original MAF-derived densities also diverge for $|z|\gtrsim 1.5$~kpc. Both these regions are regions where the number of tracer stars in our simulated galaxy is low.

In Table~\ref{tab:mass_density_at_sun}, we list the estimated mass densities at the Solar location. We see that the MAF calculation using the anisotropic kernel results in $1\sigma$ agreement with the kernel-averaged simulation-truth density.

\begin{table}
\caption{
Estimated mass densities at the Solar location, averaged over the anisotropic kernel with $\sigma_x = \sigma_y = 1$~kpc and $\sigma_z = 0.2$~kpc. We show the mass density derived from the simulation-truth, the MAF result trained on the error-free original dataset, the MAF result trained on the error-smeared dataset, and the Jeans result using the error-smeared dataset.
The origin of the different types of uncertainties are discussed in Section~\ref{subsec:errors}.
 \label{tab:mass_density_at_sun}}
\begin{center}
\begin{ruledtabular}
\begin{tabular}{lccc}
\multirow{2}{*}{Dataset} & \multicolumn{3}{c}{Mass Density $\rho$ $(10^7 \; M_{\odot} / \mathrm{kpc}^3)$} \\
& & (stat.) & (syst.)
\\
\hline
sim.-truth & $3.06$ & -- & $\pm 0.37$ \\
original & $3.33$ & $\pm 0.17$ & -- \\
smeared & $3.37$ & $\pm 0.17$ & $\pm 0.15 $\\ 
Jeans & $3.67$ &  $\pm 0.13$ & $\pm 0.05$  \\
\end{tabular}
\end{ruledtabular}
\end{center}
\end{table}

Figure~\ref{fig:density_1d} shows the estimated mass densities along $x$, $y$, and $z$ axes passing through the Solar location. 
The agreement between the simulation-truth and MAF results is best along the disk, where the vast majority of tracer stars are located. Note that the error-smeared datasets result in very similar densities as those calculated using MAFs trained on the original (error-free) dataset, indicating our bias-subtraction method (see Appendix~\ref{app:errors}) is largely accurate.

Along the disk, the largest systematic errors in our density calculations occur at large $x$ and at large positive $y$ values. The former discrepancy is due to the falling number of tracer stars at increasing distances from the galactic center, as seen in Section~\ref{sec:boltzmann}.  The deviations at large $y$ are likely due to the disequilibria within the disk, as the edge of a spiral arm is located within the observation window in this region (see the left panel of Figure~\ref{fig:equilibrium}). 

Moving off the disk to large $|z|$, the central values of our MAF-derived results largely agree out to $|z|\sim 1.5$~kpc, though again we see that the percentage errors increase as the number of tracer stars decreases dramatically. As an example of the scarcity of tracer stars at these locations, when $|z|\sim 1.5$~kpc, only $200-300$ tracer stars are within the anisotropic kernel over which we average. The actual {\it Gaia} dataset is expected to have significantly more stars even off of the disk, which may suggest that the real data could achieve more accurate results. Figure~\ref{fig:density_1d} also demonstrates that our Jeans analysis results gives significantly less accurate results with this dataset. This suggests that the MAF approach may likewise be more accurate than existing methods when applied to real datasets.

Note that the  mass density calculations have uncertainties on the order of $10\%$--$20\%$, while the errors on the magnitudes of acceleration are $\sim 1\%$--$5\%$. 
Reliable mass density calculations require very accurate phase space density fitting to reduce the noise from the  second derivatives. Our regularization scheme, including the use of GELU activation and ensemble averaging are chosen to increase the accuracy of the phase space density and its derivatives to achieve these ${\cal O}(10\%)$ errors. Given the systematic deviations in the acceleration calculations at certain points within the observation window (see Figure~\ref{fig:acceleration_1d}), it may be surprising that we can reconstruct the density accurately in regions where the accelerations were systematically off from the simulation-truth. However, the divergence calculation ignores any acceleration component which passes through the integration surface in Eq.~\eqref{eq:acceleration_integration}. This reduces the effect of the observed acceleration bias on the mass density calculation, as long as the bias itself is slowly varying though the integration surface.

\section{Conclusions \label{sec:conclusions}}

In this paper, we have demonstrated that flow-based machine learning is capable of accurately modeling the phase-space distribution -- along with first and second derivatives -- of tracer stars within realistic galaxies. The resulting distributions are accurate enough to reconstruct the acceleration field of the galaxy within $\sim 5\%$ and the mass density within $\sim 20\%$ for regions within $\sim 1.5$~kpc of the disk. This was accomplished with only ${\cal O}(10^5)$ tracer stars, which is significantly fewer than the millions of stars we may expect from the {\it Gaia} data. The impact of the relatively few tracers is more relevant away from the disk, suggesting that an analysis of real data may be more successful in measuring the acceleration and mass density even here. A larger dataset would also allow cuts on stellar properties that would mitigate effects of disequilibria in the stellar population from spiral arms within the disk. 

Compared to previous works \cite{2020arXiv201104673G,2021MNRAS.506.5721A,2021arXiv211207657N}, we have shown that flow-based modeling of the tracer phase-space density can be extended to smooth second derivatives which allows for accurate estimation of the density, as well as demonstrating the application to a fully cosmological simulation of a Milky Way-like galaxy. We do not enforce equilibrium or axisymmetry in the simulation. Realistic errors on the tracer star kinematics do not result in significant biases in the calculated accelerations and mass densities. 

The most immediate application of this machine learning approach is to measure the acceleration and density fields in real data, which will require folding in the non-uniform coverage of the sky once dust extinction is taken into account. However, the phase-space density of tracer stars is a fundamental quantity with potentially more uses beyond these. The ability of flow neural networks to accurately estimate this quantity from data may open other avenues for the study of the Galaxy's kinematics. For example, the MAF results may allow us to directly measure the departure from equilibrium in the Galaxy,  using accurate measures of acceleration which do not rely on the Boltzmann Equation, or through comparison of different subpopulations of tracers. The vast kinematic dataset made possible with {\it Gaia} is already revolutionizing our understanding of the Milky Way; unsupervised machine learning approaches such as described here have the potential to highlight and reveal additional physical information within this data.

\section*{Acknowledgements}

We thank Alyson Brooks, Sukanya Chakrabarti, David Hogg, and Adrian Price-Whelan for their advice, comments, and help.

This work has made use of data from the European Space Agency (ESA) mission {\it Gaia} (\url{https://www.cosmos.esa.int/gaia}), processed by the {\it Gaia}
Data Processing and Analysis Consortium (DPAC, \url{https://www.cosmos.esa.int/web/gaia/dpac/consortium}). Funding for the DPAC has been provided by national institutions, in particular the institutions participating in the {\it Gaia} Multilateral Agreement.

We thank the Gaia Project Scientist Support Team and DPAC for their work in development and maintenance of the \textsc{PyGaia} code.

The authors acknowledge the Office of Advanced Research Computing (OARC) at Rutgers, The State University of New Jersey for providing access to the Amarel cluster and associated research computing resources that have contributed to the results reported here. URL: \url{https://oarc.rutgers.edu}

\appendix

\section{Flow Architectures \label{app:maf}}

Our normalizing flows for estimating $\nu(\vec{x})$ and $p(\vec{v}|\vec{x})$ are constructed by using five MAFs.
Inputs to each MAF are rotated beforehand to avoid any inductive bias originating from using the same order of inputs to the masking.
For the base distribution of the flow architectures, we use the standard Gaussian.

The MADE blocks of MAFs consist of two linear layers for input and output layers and two residual blocks \cite{He_2016_CVPR} with pre-activation \cite{10.1007/978-3-319-46493-0_38} as hidden layers.
The hidden layers have 48 input features.
For these layers, we use GELU activations \cite{2016arXiv160608415H} as described in the main text.

The parameters of MADE blocks are initialized by the default scheme in \textsc{nflows}.
Most of the linear layers are initialized by a He uniform initializer \cite{He_2015_ICCV}, but only the parameters of the last linear layer of the residual block are initialized by numbers drawn from a uniform distribution on an interval [-0.001, 0.001].

\section{Preprocessor \label{app:preprocess}}

Prior to training, we preprocess both the position and velocity features of our stars drawn from \textsc{h277}. For the position, we are presented with the problem that the positions define a sphere centered on the Solar location with a sharp cut-off at $3.5$~kpc. This finite support with a sharp discontinuity is difficult to fit using MAFs which intrinsically support the entire Euclidean space and generate smooth distributions. To minimize fitting issues at the boundary in position-space, we smoothly transform the positions to the full Euclidean space $\mathbb{R}^3$ in the following way:
\begin{enumerate}
\item {\bf Centering and Scaling:} transform selected region, which is a ball centered at $\vec{x}_{\odot}$ and with radius $r_{\max}$, to a unit ball.
\begin{equation}
\vec{x} \rightarrow \frac{\vec{x} - \vec{x}_{\odot}}{r_{\max} \cdot c}
\end{equation}
where $c$ is a boundary scale multiplier, which inserts a small margin between the fiducial region of the transformed dataset and the boundary of the unit ball. In this work, we choose $c=1.000001$.

\item {\bf Radial transformation:} In order to avoid learning of sharp edges at the boundary by a continuous function,
we transform the unit ball to Euclidean space $\mathbb{R}^3$. In particular, we use the following transform and its inverse.
\begin{eqnarray}
\vec{x} 
& \rightarrow &
\vec{y}=\frac{\vec{x}}{|\vec{x}|} \tanh^{-1} |\vec{x}| 
\\
\vec{y} 
& \rightarrow &
\vec{x} = \frac{\vec{y}}{|\vec{y}|} \tanh |\vec{y}| \nonumber
\end{eqnarray}

The Jacobian determinant of this transform is as follows.
\begin{equation}
\left| \frac{d\vec{y}}{ d\vec{x}} \right| = \frac{\left( \tanh^{-1} |\vec{x}| \right)^2}{|\vec{x}|^2 (1 - |\vec{x}|^2)}
\end{equation}
The sharp edges of the window is explicitly handled by this transformation.

\item {\bf Standardization:} We scale each Cartesian direction by the standard deviation of the data along that direction:
\begin{equation}
x_i \rightarrow \frac{x_i - \langle x_i \rangle}{ \sigma_{x_i}} \, \mathrm{for} \, i=1,\cdots,3
\end{equation}
where $\langle x_i\rangle$ and $\sigma_{x_i}$ are the mean and standard deviation of the position dataset $\{\vec{x}^\alpha\}$.
\end{enumerate}

The distribution of velocities in our datasets lacks the sharp cut-off present in the positions; instead the density smoothly goes to zero at high velocities. We therefore do not need to apply a centering, scaling, or radial transformation to the $\{\vec{v}^\alpha\}$; we instead only apply a standardization step
\begin{equation}
v_i \rightarrow \frac{v_i - \langle v_i \rangle}{ \sigma_{v_i}} \, \mathrm{for} \, i=1,\cdots,3
\end{equation}

\section{Jeans Analysis \label{app:jeans}}

We wish to compare our MAF-based approach to solving the Boltzmann Equation for the accelerations with existing methods. Integrating the Boltzmann Equation weighting by velocity results in the Jeans Equation and then fitting tracer stars to these equations has been an important direct measure of local accelerations and mass densities (see Refs.~\cite{2014JPhG...41f3101R,2021RPPh...84j4901D} for detailed reviews).
Performing this analysis using the tracer star dataset from $\textsc{h277}$ therefore provides a comparison between our MAF approach and standard techniques given equivalent data.

The explicit form of Eq.~\eqref{eq:jeans} of the axisymmetric Jeans equations gives the vertical and radial acceleration fields $a_z$ and $a_R$:
\begin{eqnarray}
    a_z & = & \frac{\partial \ln(\nu)}{\partial z}\overline{v_z^2} + \frac{\partial \overline{v_z^2}}{\partial z} +  \nonumber \\
   & & \left(\frac{1}{R}+\frac{\partial \ln(\nu)}{\partial R}\right)\overline{v_R v_z} + \frac{\partial \overline{v_R v_z}}{\partial R} \label{eq:jeans-vertical} \\
    a_R & = & \frac{\partial \ln(\nu)}{\partial R}\overline{v_R^2} + \frac{\partial \overline{v_R^2}}{\partial R} + \frac{\partial \ln(\nu)}{\partial z}\overline{v_R v_z} +\nonumber \\
     & &  \frac{\partial \overline{v_R v_z}}{\partial z} + \frac{\overline{v_R^2}-\overline{v_\phi^2}}{R}. \label{eq:jeans-radial}
\end{eqnarray}

For a set of measured tracer stars, the number density $\nu$ and velocity moments $\overline{v_i v_j}$ are calculated from fits to binned data. The standard Jeans analysis literature fits these quantities to functions that work well in the Solar neighborhood of the Milky Way. These functions are not particularly well-adapted to the distribution of tracers within our observation window of $\textsc{h277}$, and so we take a data-driven approach in order to obtain the fairest possible comparison between the Jeans analysis and the MAF approach. 

Instead of using the existing functional forms, we fit the binned data to radial basis functions (RBFs). RBFs are flexible interpolators that can be fit to multidimensional data. Importantly, RBFs provide good derivative estimation across the bulk and at the boundary of the binned data. Similar non-parametric fitting methods such as the use of $B$-splines have become a subject of interest in recent Jeans analysis literature \cite{10.1093/mnras/stac400,10.1093/mnras/stx1219,10.1093/mnras/stx1798}.

We fit RBFs to data binned in cylindrical radius $R$ from the galactic center and height above the disk plane $z$ using the $\textsc{SciPy}$ {\fontfamily{qcr}\selectfont RBFinterpolator} method. $R$-bins were uniformly spaced whereas $z$-bins were spaced exponentially far apart to maintain reasonable statistics above the disk. All bins with fewer than 40 tracer stars were discarded. The increased spacing between $z$-bins reduces the accuracy of $z$-derivatives but improves fit stability. Each RBF used a multiquadric kernel with a length scale $\epsilon=3$~kpc and smoothing parameter $s=\epsilon$. In addition to enforcing axisymmetry, we impose reflection symmetry about the midplane by reflecting all stars in the observation window to $z>0$. Reflecting these tracers above the midplane doubles the number of tracers in each bin which improves their statistics, particularly for bins high above the midplane. 
One implicit consequence of reflection symmetry about the midplane is that $a_z(R,-z)=-a_z(R,z)$. For $a_z$ to be continuous across the midplane, we need to enforce $a_z(R,0)=0$. Numerically, estimates of $a_z(R,0)$ from Eq.~\eqref{eq:jeans-vertical} using our RBF fits are small, so this assumption is reasonable. However, we need to maintain perfect continuity across the midplane to be able to take reliable finite differences for estimation of $\rho$. Therefore, we offset all values of $a_z(R,z)$ at radius $R$ by the small, non-zero value of $a_z(R,0)$ to enforce $a_z(R,0)=0$. This procedure results in a slight adjustment to the value of $a_z$ at all points as well as artificially small uncertainties in $a_z$ close to the midplane (this is evident in Table~\ref{tab:acc_at_sun}) but preserves differentiability at $z=0$.

Terms in Eq.~\eqref{eq:jeans-vertical} containing $\overline{v_R v_z}$, commonly referred to as the ``tilt" terms, are neglected in many Jeans analyses as $\overline{v_R v_z}$ is small close to the midplane. As portions of our observation window reach high above the midplane, the tilt terms must be considered in our analysis. Although $\overline{v_R v_z}$ is noisy at large $|z|$, the smoothing from the RBF provides stable estimates of $\overline{v_R v_z}$ and its first few derivatives. The Jeans estimate of $a_z$ and $a_R$ compared to the MAF and truth-level accelerations are shown in Figure~\ref{fig:acceleration_1d}. The Jeans analysis successfully recovers $a_z$ and $a_R$, with the best results lying close to the center of the observation window.

We now estimate the mass density $\rho(R,z)$ from our acceleration estimates via the surface mass density $\Sigma(R,z)$, which is related to $\rho(R,z)$ via
\begin{equation}\label{eq:surface_volume_density_rel}
    \rho(R,z)=\frac{1}{2}\frac{\partial \Sigma(R,z')}{\partial z'}\biggr\rvert_z.
\end{equation}
The axisymmetric version of the Poisson equation relates $\Sigma(R,z)$ to $a_z$ and $a_R$:
\begin{equation}\label{eq:surface-density}
    \Sigma(R,z) = \frac{-1}{2 \pi G}\left(a_z(R,z) + \int_0^z \frac{1}{R} \frac{\partial (R\:a_R(R,z'))}{\partial R}dz'\right).
\end{equation}

Given our sparse binning, derivatives of acceleration higher than first-order are unstable and inaccurate. Calculations of $\rho$ via Eq.~\eqref{eq:surface_volume_density_rel} and Eq.~\eqref{eq:surface-density} involve a second-order derivative of $a_R$ that cannot be included due to its instability. Therefore, we adopt the commonly-used ``vertical approximation" for our density estimate, where $a_R$ is neglected in the calculation of $\rho$ due to its small physical contribution relative to the vertical component. The Jeans estimate of $\rho$ is shown in Figure~\ref{fig:density_1d}.

We extract two types of uncertainty for the Jeans estimate of acceleration and mass density. Uncertainties accounting for finite statistics are estimated by repeating this analysis using the 10 bootstrapped introduced in Section~\ref{subsec:errors} and calculating the standard deviation. Systematic uncertainties arising from smearing are estimated by repeating this analysis on 10 reperturbed datasets and calculating the $1\sigma$ deviation. In addition, the central value for both the acceleration and mass density has had its estimated smearing bias subtracted using the method described in Appendix~\ref{app:errors}.

\section{Acceleration Plots \label{app:accelerations}}

\begin{figure*}
\includegraphics[width=1.95\columnwidth]{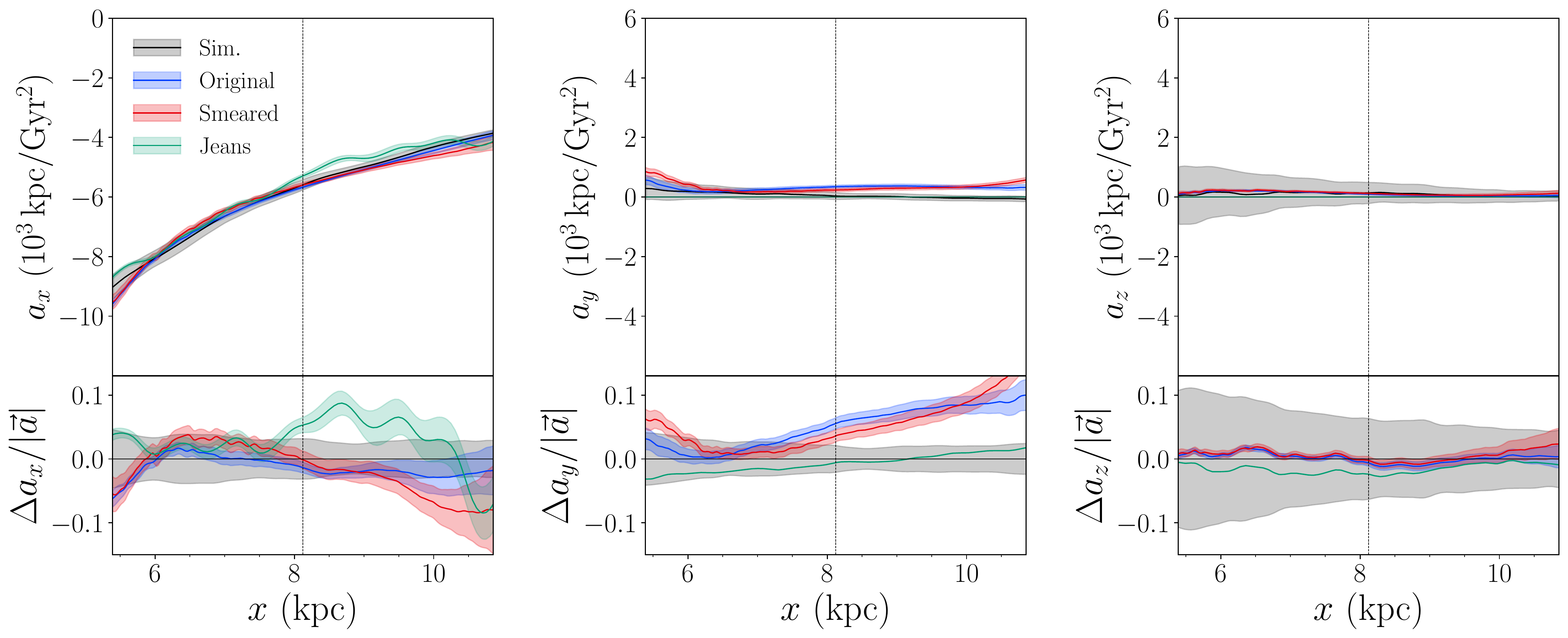}
\includegraphics[width=1.95\columnwidth]{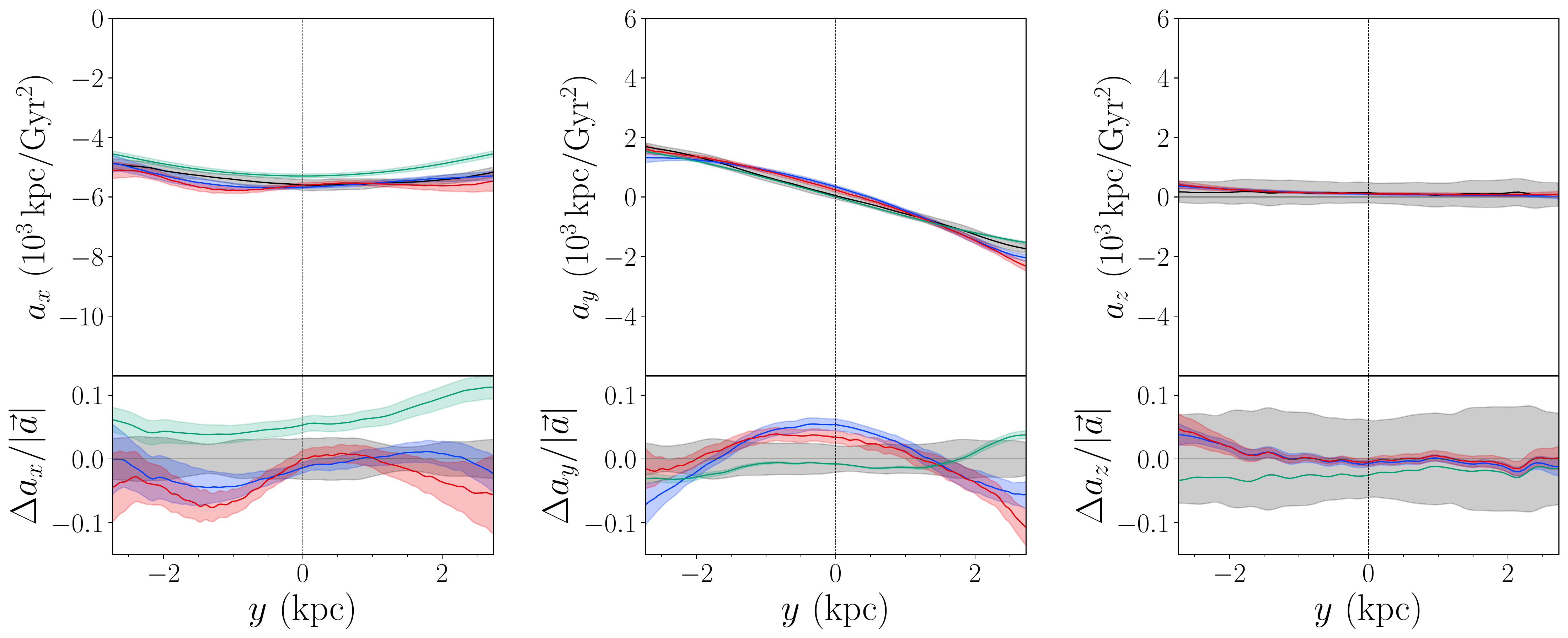}
\includegraphics[width=1.95\columnwidth]{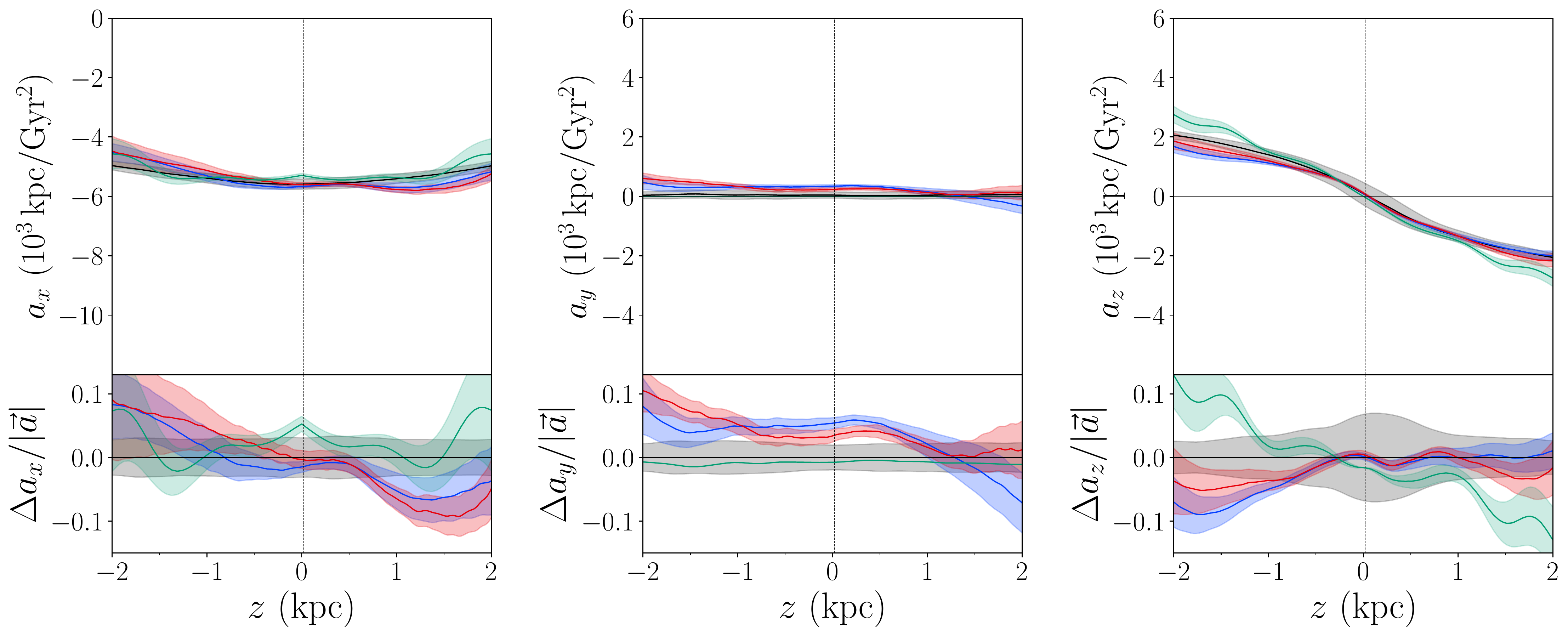}
\caption{MAF accelerations $\vec{a}$ as a function of location along the $x$-axis with $y=z=0$ (top), as a function of location along the $y$-axis with $x=8.122$~kpc and $z=0$ (middle), and as a function location along the $z$-axis  with $x=8.122$~kpc and $y=0$ (bottom) for the original dataset (blue) and the average over error-smeared dataset (red), as compared to the simulation-truth (black). The accelerations derived using the Jeans analysis approach are shown in green. Fractional deviations from the true acceleration $\Delta a_i/|\vec{a}|$ are shown below each main plot. 
The shaded regions indicate the total $1\sigma$ error estimates. \label{fig:acceleration_1d_all}}
\end{figure*}

In Section~\ref{sec:boltzmann}, we show in Figure~\ref{fig:acceleration_1d} a subset of the acceleration vectors, along three axes that pass through the Solar location. Specifically, we show $a_x$ along the $x$-axis, $a_y$ along the $y$-axis, and $a_z$ along the $z$-axis. While these three components are useful to see the agreement between the MAF calculation and the simulation-truth, calculation of the mass density requires accurate knowledge of all three components at every location, so that the volume and surface integrals can be computed. In Figure~\ref{fig:acceleration_1d_all}, we show all three components of the acceleration along each axis passing through the Solar Location.

\section{Error Subtraction \label{app:errors}}

Under the assumption of small errors drawn randomly from an error model ${\cal E}$ which is independent of position, then the smearing of the true phase space distribution $f(\vec{w})$ into the smeared distribution $f_s(\vec{w})$ is given by
\begin{equation}
f_s(\vec{w}) \equiv f\star {\cal E}(\vec{w}) = \int d^6 \vec{w}' f(\vec{w}') {\cal E}(\vec{w}-\vec{w}').
\end{equation}
A similar expression holds for the derivatives of $f$:
\begin{equation}
\frac{\partial}{\partial w_i} f_s(\vec{w}) \equiv f\star {\cal E}(\vec{w}) = \int d^6 \vec{w}' \frac{\partial}{\partial w_i} f(\vec{w}') {\cal E}(\vec{w}-\vec{w}').
\end{equation}
For small errors, we can Taylor expand the characteristic function of the error function around 
$\vec{w}' \approx \vec{w}$.
Assuming that the error model is a Gaussian with covariance matrix $\Sigma_{ij}$, the expansions to leading order in $\Sigma_{ij}$ are:
\begin{eqnarray}
f_s(\vec{w}) & \approx & f(\vec{w}) + \frac{1}{2} \frac{\partial^2 f(\vec{w})}{\partial w_i \partial w_j} \Sigma_{ij} \label{eq:f_error_bias} \\
\frac{\partial}{\partial w_i}  f_s(\vec{w}) & \approx & \frac{\partial}{\partial w_i}  f(\vec{w}) + \frac{1}{2} \frac{\partial}{\partial w_i}  \frac{\partial^2 f(\vec{w})}{\partial w_j \partial w_k} \Sigma_{jk}. \label{eq:df_error_bias}
\end{eqnarray}
This expansion clarifies what we mean by ``small'' errors: the error ellipse (encoded in $\Sigma_{ij}$) must be small compared to the scales over which the trace phase space density varies. 

The presence of measurement errors will therefore systematically bias our estimates of the phase space density and its derivatives, and therefore bias our measures of accelerations and mass densities. While it is possible to directly estimate the size of this bias in our simulated galaxy through calculation of the correction terms in Eqs.~\eqref{eq:f_error_bias} and \eqref{eq:df_error_bias} in the original and error-smeared datasets, that would not be possible in the real {\it Gaia} data, where only the error-smeared data is available. However, under the assumption of small errors and accurate knowledge of the error ellipse, we can estimate the bias from the smeared data {\it only}, and subtract it.

We take the smeared dataset and apply a random error to each tracer star according to the error model described in Section~\ref{subsec:errormodel}. It is this stage that we explicitly rely on our accurate knowledge of the error model for each tracer star $\Sigma_{ij}$, as we are assuming that our randomly-drawn errors are drawn from the same distribution as in the error-smeared dataset. For this reperturbed dataset, the phase space density and its derivatives are related to $f_s$ at leading order by
\begin{eqnarray}
f_r(\vec{w}) & \approx & f_s(\vec{w}) + \frac{1}{2} \frac{\partial^2 f_s(\vec{w})}{\partial w_i \partial w_j} \Sigma_{ij}  \nonumber  \\
 & = & f_s(\vec{w}) + \frac{1}{2} \frac{\partial^2 f(\vec{w})}{\partial w_i \partial w_j} \Sigma_{ij} +{\cal O}(\Sigma^2) \\
\frac{\partial}{\partial w_i}  f_r(\vec{w}) & \approx & \frac{\partial}{\partial w_i}  f_s(\vec{w}) + \frac{1}{2} \frac{\partial}{\partial w_i}  \frac{\partial^2 f_s(\vec{w})}{\partial w_j \partial w_k} \Sigma_{jk} \nonumber \\
& = &  \frac{\partial}{\partial w_i}  f_s(\vec{w}) + \frac{1}{2} \frac{\partial}{\partial w_i}  \frac{\partial^2 f(\vec{w})}{\partial w_j \partial w_k} \Sigma_{jk} +{\cal O}(\Sigma^2). \nonumber
\end{eqnarray}
That is, to leading order in $\Sigma$, the bias between the reperturbed dataset and the smeared dataset is the same as the bias between the smeared dataset and the original, error-free data. This justifies our approach to subtracting the measurement-error bias by generating reperturbed datasets, training new MAFs, and calculating the average shift in derived properties (e.g., accelerations and mass densities), which can then be subtracted off the derived quantities calculated using the smeared datasets.

\bibliography{density}

\end{document}